%% file: camera-ready.tex
\documentclass[a4paper,UKenglish,cleveref, autoref, thm-restate]{lipics-v2021}
\pdfoutput=1 %uncomment to ensure pdflatex processing (mandatatory e.g. to submit to arXiv)
\hideLIPIcs  %uncomment to remove references to LIPIcs series (logo, DOI, ...), e.g. when preparing a pre-final version to be uploaded to arXiv or another public repository

\graphicspath{{./figures/}}%helpful if your graphic files are in another directory

\title{Node Labeling in Line Diagrams of Ordered Sets}

\author{Marcel Nöhre}
{
    Knowledge \& Data Engineering Group, University of Kassel, Germany \and
    Interdisciplinary Research Center for Information System Design, University of Kassel, Germany \and 
    \url{https://www.kde.cs.uni-kassel.de/noehre} 
}
{noehre@cs.uni-kassel.de}
{https://orcid.org/0009-0005-4089-2925}
{} % funding acknowledgements

\author{Gerd Stumme}
{
    Knowledge \& Data Engineering Group, University of Kassel, Germany \and
    Interdisciplinary Research Center for Information System Design, University of Kassel, Germany \and 
    \url{https://www.kde.cs.uni-kassel.de/stumme} 
}
{stumme@cs.uni-kassel.de}
{https://orcid.org/0000-0002-0570-7908}
{} % funding acknowledgements

\authorrunning{M. Nöhre and G. Stumme}

\Copyright{Marcel Nöhre and Gerd Stumme}

\ccsdesc[500]{Mathematics of computing~Graph algorithms}
\ccsdesc[500]{Human-centered computing~Graph drawings}

\keywords{Node Labeling, Line Diagrams, Formal Concept Analysis}

\category{Regular Paper} %optional, e.g. invited paper

\relatedversion{} %optional, e.g. full version hosted on arXiv, HAL, or other respository/website
\supplement{Software}%optional, e.g. related research data, source code, ... hosted on a repository like zenodo, figshare, GitHub, ...
\supplementdetails[subcategory={Source Code}]{Software}{https://github.com/marcelnoehre/node-labeling}

\acknowledgements{We would like to thank the anonymous reviewers for their comments and suggestions, which helped improve the quality of this work.}

\nolinenumbers %uncomment to disable line numbering

\EventEditors{John Q. Open and Joan R. Access}
\EventNoEds{2}
\EventLongTitle{42nd Conference on Very Important Topics (CVIT 2016)}
\EventShortTitle{CVIT 2016}
\EventAcronym{CVIT}
\EventYear{2016}
\EventDate{December 24--27, 2016}
\EventLocation{Little Whinging, United Kingdom}
\EventLogo{}
\SeriesVolume{42}
\ArticleNo{23}
\usepackage[utf8]{inputenc}
\usepackage[T1]{fontenc}
\usepackage{graphicx}
\usepackage{graphics}
\usepackage{fca}
\usepackage{tikz}
\usetikzlibrary{shapes.symbols, shapes.geometric, arrows.meta, positioning, fit, backgrounds}
\usepackage{tikzscale}
\newcommand{\old}[1]{{\color{teal} #1}}

\renewcommand{\old}[1]{ }
\newcommand{\commentout}[1]{ }
\newcommand{\R}{\mathbb{R}}
\newcommand{\ie}{i.\,e.\xspace}

\newcommand{\eg}{e.\,g.\xspace}
\newcommand{\K}{\mathbb{K}}

\DeclareMathOperator{\POS}{pos}

\DeclareMathOperator{\CH}{CH}

\DeclareMathOperator*{\argmin}{arg\,min}
\newcommand{\RR}{\mathcal{R}}

\newcommand{\CC}{\mathcal{C}}
\newcommand{\LL}{\mathcal{L}}

\newcommand{\HH}{\mathcal{H}}
\newcommand{\OO}{\mathcal{O}}
\newcommand{\PP}{\mathcal{P}}
\renewcommand{\SS}{\mathcal{S}}
\renewcommand{\AA}{\mathcal{A}}

\let\cref\Cref

\begin{document}

\maketitle

\begin{abstract}
    We propose a flexible, two-phase algorithm for labeling line diagrams of ordered sets, in which the nodes of direct neighbors in the order relation are connected by a straight, upward-pointing line. In contrast to the labeling of diagrams of arbitrary graphs, we benefit from the fact that all edges in line diagrams of ordered sets are more or less vertical. In this paper, we study the placement of all labels such that they do not intersect with any nodes, lines, or other labels while minimizing the distances between the nodes and their labels.

    Our approach starts by filtering the fixed-position model using line diagram-specific readability criteria. For labels that cannot be placed adjacent to their node (overflow labels), we exploit the free space in the graph's interior or the infinite space surrounding the drawing and link the labels with their respective node by straight binding lines that should not cross other nodes or labels if possible. To balance quality and runtime, we derive an initial placement of the overflow labels using a cost function over a sparse grid of candidates, followed by a force-based refinement step to fine-tune the layout. 

    Furthermore, we demonstrate the flexibility of this approach by applying it to the visual constraints of line diagrams in the field of Formal Concept Analysis (FCA), where certain labels have to be placed above their node and others below. This special version of the algorithm shows that a pre-filtering in the first phase and minimal adjustments for the cost function and force-based model are sufficient to handle the dual labeling requirements of concept lattices.

    \subparagraph{Generative AI Declaration}
    Generative AI was not used in the preparation of this article.
\end{abstract}

%%%%%%%%%%%%%%%%%%%%%%%%%%%%%%%%%%%%%%%%%%%%%%%%%%%%%%%%%%%%%%%%%%%%%%%%%%%%%%%%
\section{Introduction}
\label{sec:introduction}
%%%%%%%%%%%%%%%%%%%%%%%%%%%%%%%%%%%%%%%%%%%%%%%%%%%%%%%%%%%%%%%%%%%%%%%%%%%%%%%%
Order theory is a mathematical branch that deals with the modeling of hierarchies and temporal relationships. Applications include ontology engineering, scheduling problems, bibliometrics, the semantics of programming languages, and version control systems. In terms of graph theory, an ordered set is a directed, transitive, acyclic graph. Ordered sets are commonly visualised in so-called line diagrams, which only show the transitive reduction of the edges. These edges are drawn as straight, upward-directed lines. Several algorithms for drawing such line diagrams exist, \eg~\cite{sugiyama1981methods, freese2004automated, Duerrschnabel2023, Noehre2026}.

These algorithms focus on node placement, and the labels are then attached at a fixed offset above or below each node. Surprisingly little work exists on the labeling itself. Automated labeling of graph diagrams in general is NP-hard~\cite{Formann1991} and has been studied extensively in fields like cartography~\cite{Doddi1998, Stein2008}, where models such as the fixed-position model and the slider model place each label anywhere in the free space around its point~\cite{Formann1991, Kreveld1999}. However, we focus on a more specific task, an upward-directed line diagram with no edges close to horizontal, so these methods alone are not suitable. Moreover, we consider labels whose position relative to the node is meaningful.

Given a fixed line diagram of an ordered set and a corresponding set of \LaTeX\ labels, we address the problem of automatically placing each label horizontally as close as possible to its assigned node. This layout must be achieved by respecting the community standards for labeling line diagrams of ordered sets: no label may intersect any node, edge, or other label, label scaling is strictly prohibited, and the line diagram itself may not be modified, since it usually emphasizes structural properties of the ordered set (such as the order dimension). We target this problem in the setting of static diagram generation for publication, where a line diagram is finalized once and rendered as a standalone figure, \eg for a book, paper, or poster. This is different from interactive visualization systems for large diagrams~\cite{Pattison2014}, where labels can be filtered or revealed on demand as the user explores the graph; our static setting instead requires full label coverage rather than a best-effort subset.

\begin{figure}[t]
	\centering
	$\vcenter{\hbox{\includegraphics[width=0.45\textwidth]{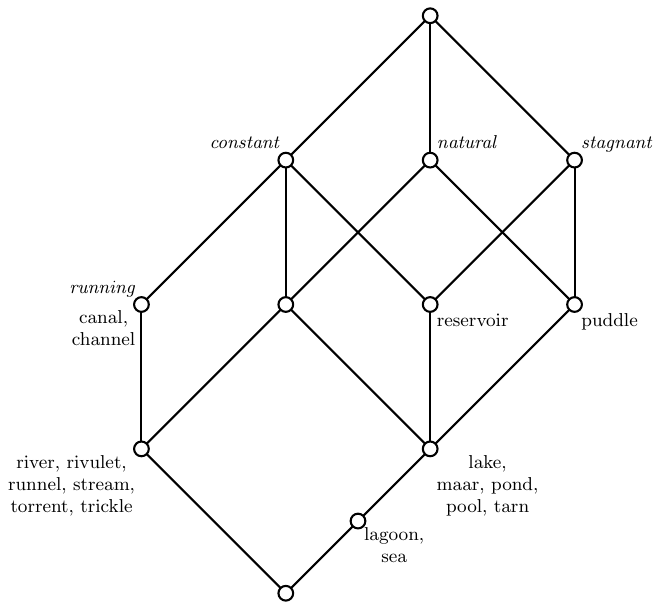}}}$%
	$\vcenter{\hbox{\includegraphics[width=0.55\textwidth]{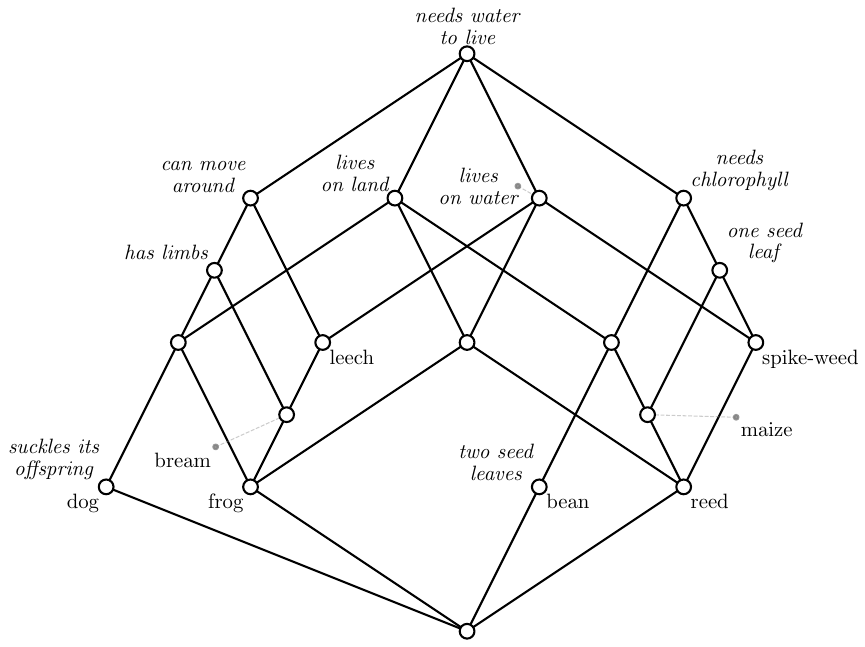}}}$%
	\vspace{1em}
	$\vcenter{\hbox{\includegraphics[width=0.45\textwidth]{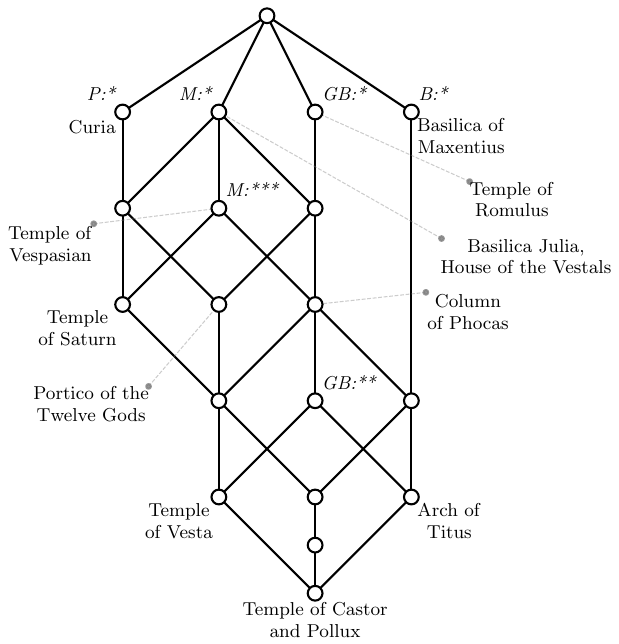}}}$%
	$\vcenter{\hbox{\includegraphics[width=0.55\textwidth]{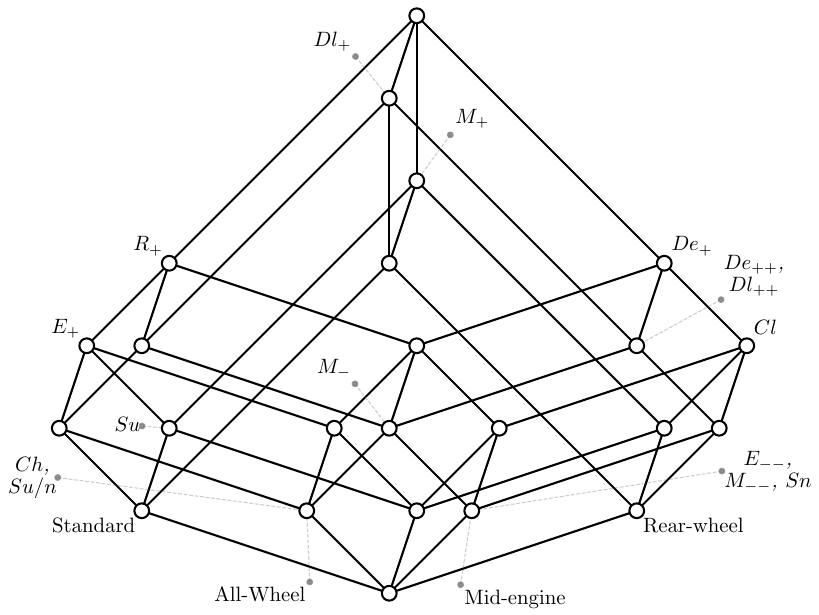}}}$%
	\caption{Automated labeling in line diagrams of ordered sets using our proposed algorithm.}
	\label{fig:evaluation_real_world}
\end{figure}

We consider two such tasks. The first covers arbitrary ordered sets and arbitrary collections of labels. In the second, we turn to Formal Concept Analysis~\cite{GanterFCA2024}, where the order carries more structure: a formal context of objects, attributes, and an incidence between them induces a concept lattice. Here the labels are not arbitrary. Each object sits at the smallest concept whose extent contains it, drawn below the node, and each attribute at the largest whose intent contains it, drawn above. The diagram can then be read directly, downward from a node for its extent and upward for its intent. Typical outcomes of the second task are shown in Figure~\ref{fig:evaluation_real_world}.

We propose a two-phase approach that balances local precision with a global heuristic to ensure full coverage. In the first phase, we begin by generating a set of label candidates based on a fixed-position model~\cite{Christensen1995}, which is iteratively filtered against the specific geometric constraints of line diagrams. Remaining conflicts are resolved by an adapted hybrid map-labeling algorithm~\cite{Wagner1998} that identifies the maximal conflict-free subset of labels, further refined by a readability tie-breaker to prioritize the most intuitive placements.

In the second phase, we address the remaining unplaced labels --- so-called overflow labels --- which cannot be positioned directly adjacent to their node. Instead, we place them within the nearest sufficient graph face or the infinite exterior surrounding the drawing. To maintain a clear association, these labels are connected to their respective nodes via straight binding lines, which are not allowed to intersect other nodes and should avoid intersecting other labels if possible. Furthermore, we aim to minimize intersections with edges or other binding lines. To this end, we determine an initial placement using an iterative Hungarian solver, followed by a force-based refinement phase to fine-tune the final layout.

The paper is organized as follows: after recalling the basics of ordered sets, lattices, planar graphs, and spatial data structures in Section~\ref{sec:preliminaries}, we state the problem to be solved in Section~\ref{sec:problem} and introduce our node labeling algorithm for line diagrams of ordered sets (Section~\ref{sec:algorithm}). Furthermore, we provide a specialized version for concept lattices in Section~\ref{sec:cl_algorithm} that manages the dual requirements of intent and extent labels, placing them strictly above and below their respective nodes to respect visual conventions. Finally, in Section~\ref{sec:evaluation}, we present a runtime analysis to derive a default parameterization that balances computational effort and visual accuracy, before concluding with directions for future work in Section~\ref{sec:conclusion}.

%%%%%%%%%%%%%%%%%%%%%%%%%%%%%%%%%%%%%%%%%%%%%%%%%%%%%%%%%%%%%%%%%%%%%%%%%%%%%%%%
\section{Preliminaries}
\label{sec:preliminaries}
%%%%%%%%%%%%%%%%%%%%%%%%%%%%%%%%%%%%%%%%%%%%%%%%%%%%%%%%%%%%%%%%%%%%%%%%%%%%%%%%
In this section, we recall the basics of ordered sets, lattices~\cite{Birkhoff1967, Davey2002}, and planar graphs~\cite{Diestel2017}, followed by an introduction to minimal bounding rectangles~\cite{Freeman1975} for spatial indexing of geometric intersections~\cite{Leutenegger1997}.

%%%%%%%%%%%%%%%%%%%%%%%%%%%%%%%%%%%%%%%%%%%%%%%%%%%%%%%%%%%%%%%%%%%%%%%%%%%%%%%%
\subsection{Ordered sets}
\label{subsec:ordered_sets}
%%%%%%%%%%%%%%%%%%%%%%%%%%%%%%%%%%%%%%%%%%%%%%%%%%%%%%%%%%%%%%%%%%%%%%%%%%%%%%%%
An \emph{ordered set} is a pair $(P, \leq)$, where $P$ is a set and $\leq$ is a binary relation satisfying reflexivity ($p \leq p$), antisymmetry ($p \leq q \text{ and } q \leq P \implies p = q$), and transitivity ($p \leq q \text{ and } q \leq z \implies p \leq z$). The relation $\leq$ determines the comparability of elements; specifically, $p, q \in P$ are \emph{comparable} if $p \leq q$ or $q \leq p$, otherwise they are \emph{incomparable} $p \parallel q$. If every pair of elements in $P$ is comparable, the set is called a \emph{totally ordered set} or a \emph{chain}. The \emph{covering relation} $\prec$ of a poset is defined such that $p \prec q$ if $p < q$ and there is no element $z \in P$ with $p < z < q$ (in other words: $\prec$ is the transitive-reflexive reduction of $\leq$). We then call $p$ a \emph{lower cover} of $q$, and $q$ an \emph{upper cover} of $p$. A \emph{lattice} $L$ is a special case of an ordered set in which every pair of elements $\{p, q\} \in L$ has a unique least upper bound, denoted as their \emph{join} ($p \vee q$), and a unique greatest lower bound, denoted as their \emph{meet} ($p \wedge q$).

A \emph{line diagram} $\Gamma$ of an ordered set $(P, \leq)$ is the directed graph $(V, E) := (P, \prec)$ together with a placement function $\POS \colon V \to \R^2$, which fulfills the condition that $p \prec q$ implies $\POS(p)_y < \POS(q)_y$. By convention, all edges are drawn as straight lines. Because they are all upward-directed, the arrowheads are omitted.

%%%%%%%%%%%%%%%%%%%%%%%%%%%%%%%%%%%%%%%%%%%%%%%%%%%%%%%%%%%%%%%%%%%%%%%%%%%%%%%%
\subsection{Planar graphs}
\label{subsec:planar}
%%%%%%%%%%%%%%%%%%%%%%%%%%%%%%%%%%%%%%%%%%%%%%%%%%%%%%%%%%%%%%%%%%%%%%%%%%%%%%%%
A graph $G = (V, E)$ is \emph{planar} if it can be drawn on the Euclidean plane $\R^2$ without edge crossings. Such a partition yields \emph{faces} $F$ corresponding to one-dimensional holes and one unbounded exterior region, satisfying Euler's formula ($|V| - |E| + |F| = 2$). Given any two-dimensional drawing $\Gamma$ of a non-planar graph $G = (V, E)$, \emph{planarization} is the process of transforming it into a planar graph $G' = (V', E', F')$, by treating every edge crossing as a structural vertex. Inserting a dummy vertex at each intersection and subdividing the intersecting edges yields a strictly planar graph, with a set of faces $F'$ that corresponds to the distinct visual holes created by the intersecting paths in the original drawing.

A \emph{walk} in a graph $G = (V, E)$ is a sequence of vertices $(v_0, \ldots, v_k)$ such that $\{v_{i-1}, v_i\} \in E$ for all $1 \leq i \leq k$. For a planar embedding, the \emph{boundary} of a face $f \in F$, denoted by $\partial f$, is the closed walk of boundary edges that encloses the region. In a planarized graph  $G'$, each face $f' \in F'$ is bounded by a cycle of vertices $v' \in V'$, where $V'$ includes both the original nodes and the dummy vertices corresponding to edge crossings. The \emph{unbounded face} $f'_\infty$ is the infinite region in $\R^2 \setminus \Gamma$ that is defined by its boundary walk $\partial f'_\infty$. As the boundary of the unbounded face may be non-convex, we consider the \emph{concave hull} $\CH(G)$, which is the polygon defined by the boundary walk $\partial f'_\infty$ enclosing all peripheral nodes.

%%%%%%%%%%%%%%%%%%%%%%%%%%%%%%%%%%%%%%%%%%%%%%%%%%%%%%%%%%%%%%%%%%%%%%%%%%%%%%%%
\subsection{Spatial Data Structures}
\label{subsec:spatial}
%%%%%%%%%%%%%%%%%%%%%%%%%%%%%%%%%%%%%%%%%%%%%%%%%%%%%%%%%%%%%%%%%%%%%%%%%%%%%%%%
To handle label intersections efficiently, we utilize minimum bounding rectangles and R-trees for spatial indexing~\cite{Freeman1975}. A \emph{minimum bounding rectangle} (MBR), denoted as $\RR(o)$, is the smallest axis-aligned rectangle enclosing a spatial object $o \subseteq \R^2$. Formally, for an arbitrary two-dimensional object $o$ defined by a set of points $\{x, y\}$, the MBR is the Cartesian product of the intervals spanning its projections onto the $x$ and $y$ axes:
\[
\RR(o) = [\operatorname{min}(x), \operatorname{max}(x)] \times [\operatorname{min}(y), \operatorname{max}(y)],
\]
which simplifies arbitrary polygons to a 4-tuple ($x_{\operatorname{min}}, x_{\operatorname{max}}, y_{\operatorname{min}}, y_{\operatorname{max}}$), thus enabling efficient spatial indexing and overlap detection through simple coordinate comparisons.

To manage spatial data efficiently, an \emph{R-tree} partitions $\R^d$ into a hierarchy of MBRs~\cite{Guttman1984}, ensuring a logarithmic height of $O(\log_m N)$ by maintaining between $m$ and $M$ entries per node. Therefore, the efficiency of the structure is highly sensitive to how much these MBRs overlap ($\RR_i \cap \RR_j \not= \emptyset$)~\cite{Orenstein1986}. For static datasets, such as line diagrams of ordered sets, the Sort-Tile-Recursive (STR) algorithm further improves the performance by using a multi-dimensional packing scheme~\cite{Leutenegger1997}.

%%%%%%%%%%%%%%%%%%%%%%%%%%%%%%%%%%%%%%%%%%%%%%%%%%%%%%%%%%%%%%%%%%%%%%%%%%%%%%%%
\section{Problem Definition}
\label{sec:problem}
%%%%%%%%%%%%%%%%%%%%%%%%%%%%%%%%%%%%%%%%%%%%%%%%%%%%%%%%%%%%%%%%%%%%%%%%%%%%%%%%
Given a fixed line diagram $\Gamma$ of an ordered set $(P, \leq)$, and a set $\LL$ of labels that are assigned to elements of $P$ via a function $\alpha \colon \LL \to P$. The labels are arbitrary strings, which in our case are rendered \LaTeX\ expressions.

For each label $\ell \in \LL$, we seek a placement $\Psi(\ell) := (p, a, w, h, \sigma)$, where $p := \alpha(\ell)$ is the node assigned to label $\ell$, $a \in \AA$ is the chosen anchor position from the 8-position model, $w$ and $h$ are the width and height of the label's minimal bounding rectangle, and $\sigma$ is an offset vector, which is visually represented as a binding line from the anchor point to the respective node.  

We formulate label placement as a constrained optimization problem. A placement $\Psi$ is valid if it satisfies two hard constraints: (1) the label must not intersect any other label, node, or edge in the line diagram, and (2) if $\sigma \neq (0, 0)$, the binding line must not intersect any other label or unrelated node. Among all valid placements, we seek the one that locally minimizes the total offset
\[
\min_{\sigma} \; \sum_{\ell \in \mathcal{L}} \|\sigma_\ell\| \;-\; c \cdot \sum_{\ell \in \mathcal{L}} \min_{o \in \OO} d(\sigma_\ell, o),
\]
for a given weighting factor $c \in \R^+$ balancing offset minimization against obstacle clearance, to ensure that labels do not cluster too closely together.

%%%%%%%%%%%%%%%%%%%%%%%%%%%%%%%%%%%%%%%%%%%%%%%%%%%%%%%%%%%%%%%%%%%%%%%%%%%%%%%%
\section{General Algorithm}
\label{sec:algorithm}
%%%%%%%%%%%%%%%%%%%%%%%%%%%%%%%%%%%%%%%%%%%%%%%%%%%%%%%%%%%%%%%%%%%%%%%%%%%%%%%%
In a first phase, we derive a set of candidate placements for each label based on the 8-position model, which defines potential anchor positions attached to the node. From this, we derive the maximum subset of single best candidates that can be placed without intersecting nodes, lines, or other labels. For the remaining unplaced labels, we first attempt to find valid placements within the bounded faces of the graph and then apply a global heuristic to find placements in the unbounded face, using a Hungarian solver and a force-based refinement. 

\begin{figure}[t]
  \centering
  \raisebox{1ex}{\includegraphics[width=0.4625\textwidth]{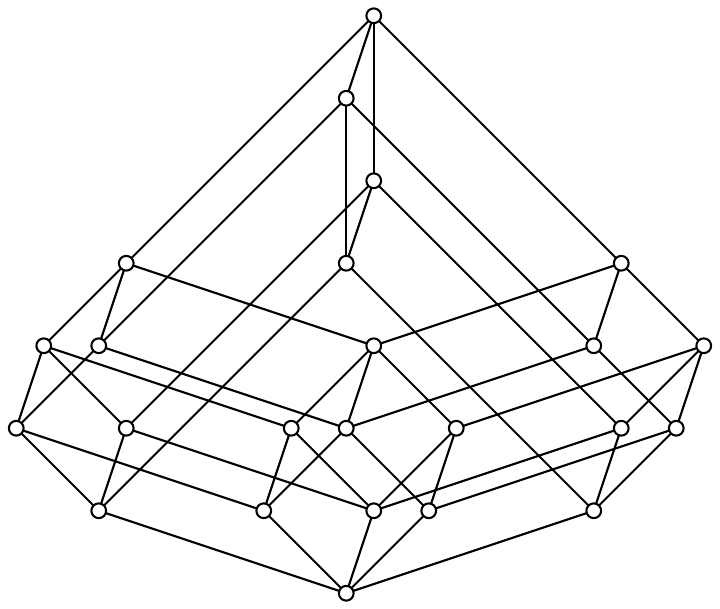}}%
  \hfill
  \includegraphics[width=0.5375\textwidth]{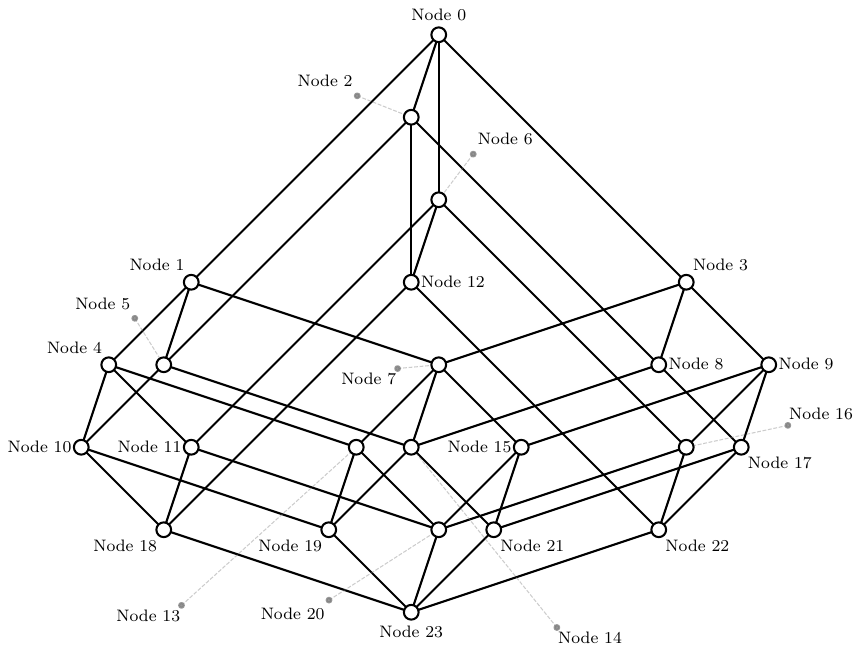}%
  \caption{Original drawing $\Gamma$ of $G$ (left) and the labeled version using our algorithm (right).}
  \label{fig:original_input}
\end{figure}

As running example consider a hand-drawn version of the lattice of driving concepts (Figure~\ref{fig:original_input}), where each node shall be labeled with a unique identifier in the format ``Node \textit{index}''. In section~\ref{sec:cl_algorithm}, we will explain the origin of this lattice in more detail.

\begin{figure}[t]
  \centering
  \includegraphics[width=0.43\textwidth]{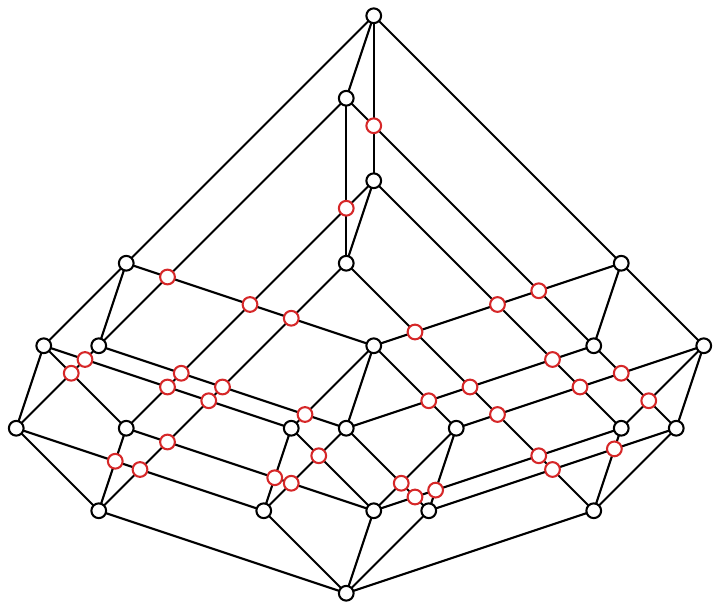}%
  \hfill
  \includegraphics[width=0.5425\textwidth]{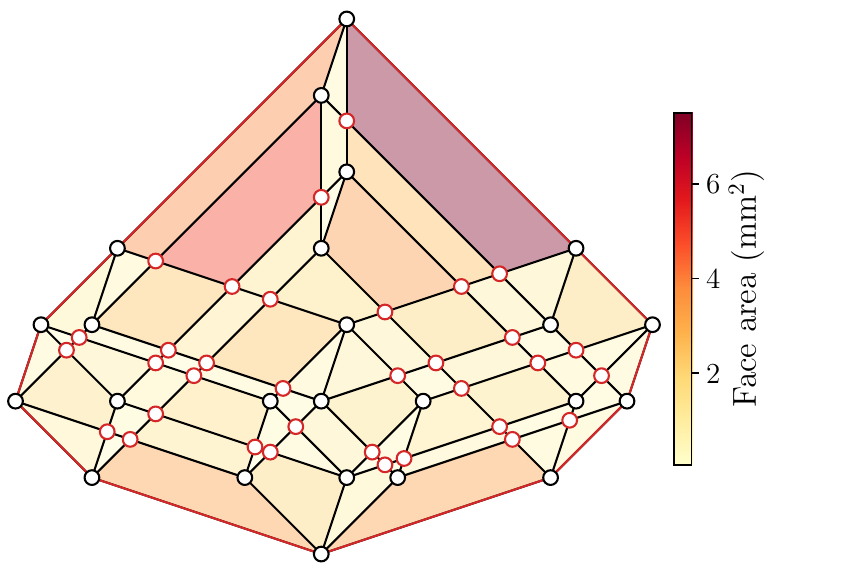}%
  \caption{The planarized graph $G'$ with dummy vertices (left) and face sizes of $F'$ (right).}
  \label{fig:planarize_input}
\end{figure}

To enable more efficient computations, we begin by transforming the graph $G = (V, E)$ into a planar representation (Figure~\ref{fig:planarize_input}) by identifying all edge crossings using the STR algorithm. To optimize this search, we exploit structural invariants~\cite{Brinkhoff1993} to filter out MBR intersections caused by edges sharing a common vertex $v \in V$. By inserting dummy vertices at the geometric intersection points, we derive an augmented planar graph $G' = (V', E', F')$ that allows us to exploit topological properties of faces $F'$, enabling more efficient computations in the subsequent steps.

\begin{figure}[t]
  \centering
  \input{figures/algorithm}
  \caption{Overview of our label placement approach. Each step is referenced with its respective section identifier.}
  \label{fig:algorithm}
\end{figure}
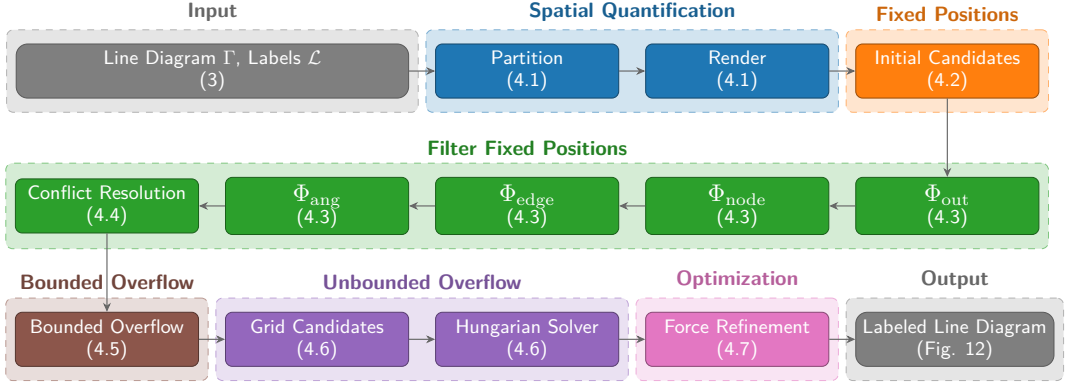

%%%%%%%%%%%%%%%%%%%%%%%%%%%%%%%%%%%%%%%%%%%%%%%%%%%%%%%%%%%%%%%%%%%%%%%%%%%%%%%%
\subsection{Label formatting and spatial quantification}
\label{subsec:label}
%%%%%%%%%%%%%%%%%%%%%%%%%%%%%%%%%%%%%%%%%%%%%%%%%%%%%%%%%%%%%%%%%%%%%%%%%%%%%%%%
Each label $\ell \in \LL$ consists of an ordered sequence of words $W = \{ w_1, w_2, \ldots, w_n \}$, where each word $w_j \in W$ is an atomic, indivisible unit. This is particularly relevant for \LaTeX\ sequences enclosed in math delimiters (\eg, \$$\ldots$\$), as it prevents inappropriate line breaks within mathematical expressions. To achieve a balanced visual presentation, we employ the Knuth-Plass algorithm~\cite{Knuth1981} to partition $W$ into a vertical stack of $k$ rows. Therefore, we derive the rendered width $\|w_j\|$ for each word, based on a specific typographic configuration (font size, typeface, etc.). The algorithm then identifies the line breaks that minimize the variance $\sum_{i = 1}^k (\tau - \|R_i\|)^2$ of row widths relative to a target width $\tau$, where $\|R_i\|$ represents the total rendered width of all words in row $i$. Finally, the partitioned label is treated as a single spatial object, which is then quantified by the minimum bounding rectangle $\RR(\ell)$ capturing all rendered glyphs. The label's spatial extent is thus defined by width $w$ and height $h$ of the rectangle $\RR(\ell)$, which are used in subsequent placement computations $\Psi(\ell) := (p, a, w, h, \sigma)$.

%%%%%%%%%%%%%%%%%%%%%%%%%%%%%%%%%%%%%%%%%%%%%%%%%%%%%%%%%%%%%%%%%%%%%%%%%%%%%%%%
\subsection{Fixed-position candidates}
\label{subsec:fixed_position}
\begin{figure}[t]
    \centering
    \hspace*{\fill}
    $\vcenter{\hbox{\includegraphics[width=0.25\textwidth]{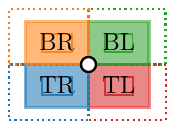}}}$%
    \hspace*{4em}%
    $\vcenter{\hbox{\includegraphics[width=0.25\textwidth]{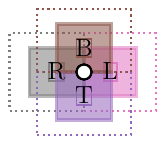}}}$%
    \hspace*{\fill}
    \caption{Label candidates $\Lambda_\ell$ based on the 8-position model, showing the corner-aligned anchors (left) and the side-aligned anchors (right).}
    \label{fig:candidate_anchors}
\end{figure}

%%%%%%%%%%%%%%%%%%%%%%%%%%%%%%%%%%%%%%%%%%%%%%%%%%%%%%%%%%%%%%%%%%%%%%%%%%%%%%%%
For each label $\ell \in \LL$ assigned to a node $v \in V$, we generate a set of eight candidate label positions $\Lambda_\ell$ based on the 8-position model~\cite{Christensen1995}. Given the minimal bounding rectangle $\RR(\ell)$ of the label's glyphs, we apply a padding $\delta \in \R^+$ to define the bounding box $\RR_{\delta}(\ell)$, which serves as the visual space for the label. Each candidate $\lambda_j \in \Lambda_\ell$ is anchored at one of the corners or midpoints of this bounding box. Furthermore, we define an extended padding $\varepsilon$ towards all sides of the bounding box that are not touched by the anchoring point, which yields the extended bounding box $\RR_{\varepsilon}(\ell)$ that is used to ensure a clear association between the label and its corresponding node (Figure~\ref{fig:candidate_anchors}).

\begin{figure}[t]
    \centering
    \includegraphics[width=0.5\textwidth]{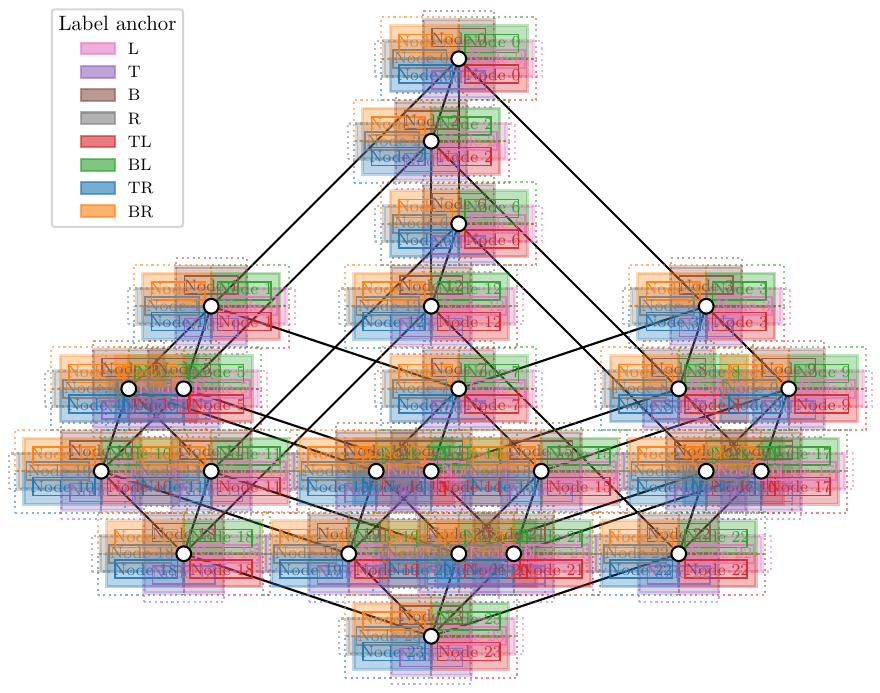}%
    \caption{Initial set of candidates $\Lambda$ for the lattice of driving concepts.}
    \label{fig:all_candidates}
\end{figure}

Revisiting our example from Figure~\ref{fig:original_input}, we assign the eight label candidates $\Lambda_\ell$, to each node $v \in V$ (Figure~\ref{fig:all_candidates}). For subsequent computations, each label candidate $\lambda \in \Lambda_\ell$ is fully defined by the position of its respective node, the anchoring point, and the size of the three axis-aligned bounding rectangles $\{\RR, \RR_\delta, \RR_\varepsilon\}$. The binding line is trivial for this phase given by $\sigma = (0, 0)$, as the label is directly anchored to the node.

%%%%%%%%%%%%%%%%%%%%%%%%%%%%%%%%%%%%%%%%%%%%%%%%%%%%%%%%%%%%%%%%%%%%%%%%%%%%%%%%
\subsection{Local geometric constraints}
\label{subsec:local_filter}
%%%%%%%%%%%%%%%%%%%%%%%%%%%%%%%%%%%%%%%%%%%%%%%%%%%%%%%%%%%%%%%%%%%%%%%%%%%%%%%%
Given the candidate set $\Lambda$, we apply a series of local geometric filters, sorted by increasing computational complexity, to prune candidates that violate the previously defined hard constraints for labeling line diagrams of ordered sets. Furthermore, we already exclude candidates in dense regions if a non-empty subset of better-positioned candidates remains.

The first filter $\Phi_{\text{out}}$ restricts the candidate set to positions that lie outside the concave hull $\CH(G)$, thereby preserving the visual integrity of the graph's interior. For any label $\ell$ assigned to a node $v \in V$ on the boundary of the concave hull $\partial f'_\infty$, the filter $\Phi_{\text{out}}(\ell)$ retains only those candidates $\lambda$ whose minimal bounding rectangle $\RR(\lambda)$ does not intersect the interior of the concave hull:
\[
\Phi_{\text{out}}(\ell) = 
\begin{cases}
    \{\lambda \in \Lambda_\ell \mid \RR(\lambda) \cap \CH(G) = \emptyset \} & \text{if } v \in \partial f'_\infty \\
    \Lambda_\ell & \text{otherwise.}
\end{cases}
\]

\begin{figure}[t]
    \centering
    \includegraphics[width=0.45\textwidth]{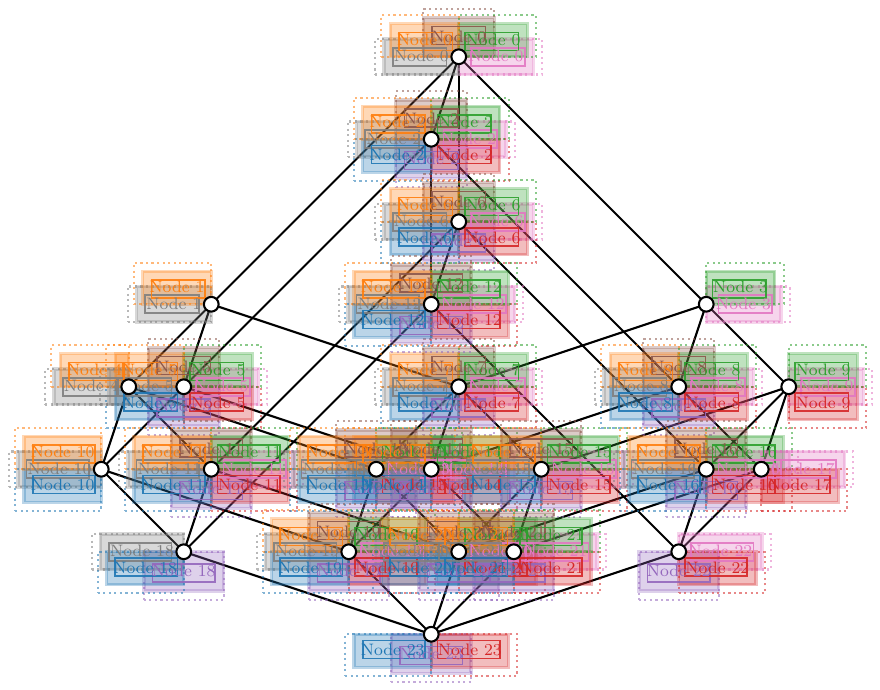}
    \hfill
    \includegraphics[width=0.45\textwidth]{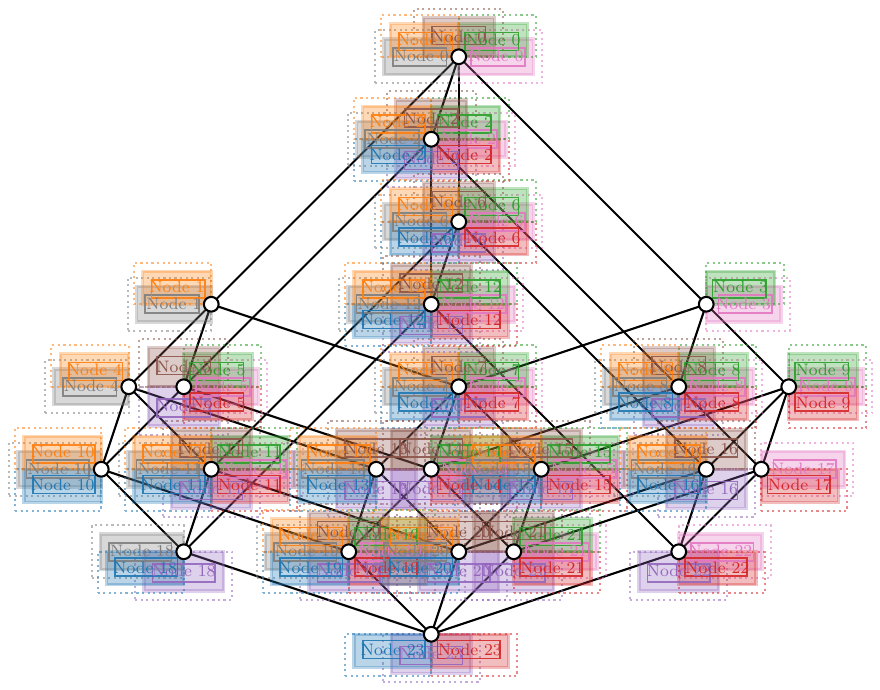}
    \caption{Candidates filtered by outer nodes (left) and unclear node assignment (right).}
    \label{fig:filtered_first}
\end{figure}

\begin{figure}[t]
    \centering
    \includegraphics[width=0.45\textwidth]{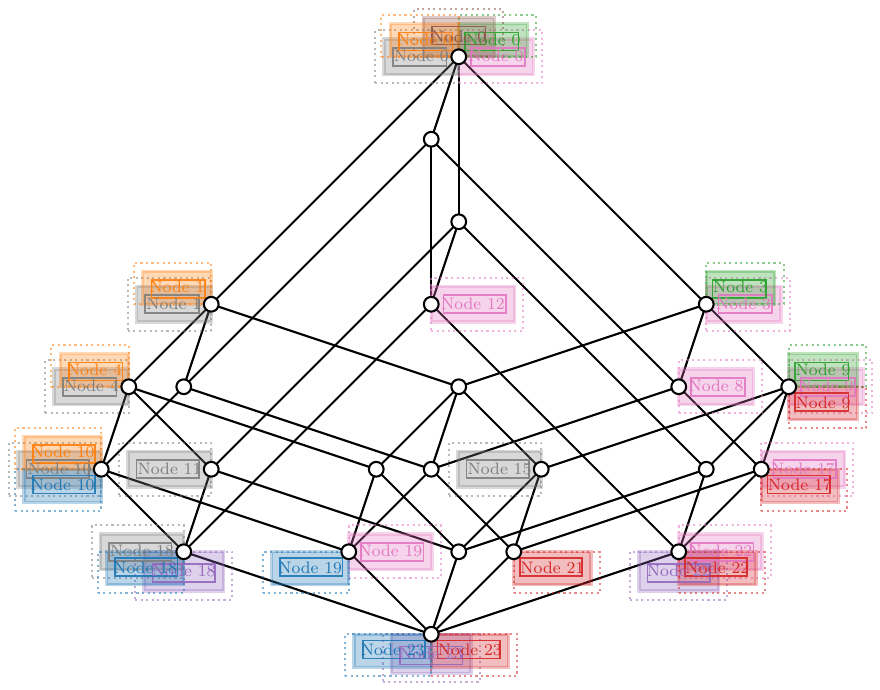}
    \hfill
    \includegraphics[width=0.45\textwidth]{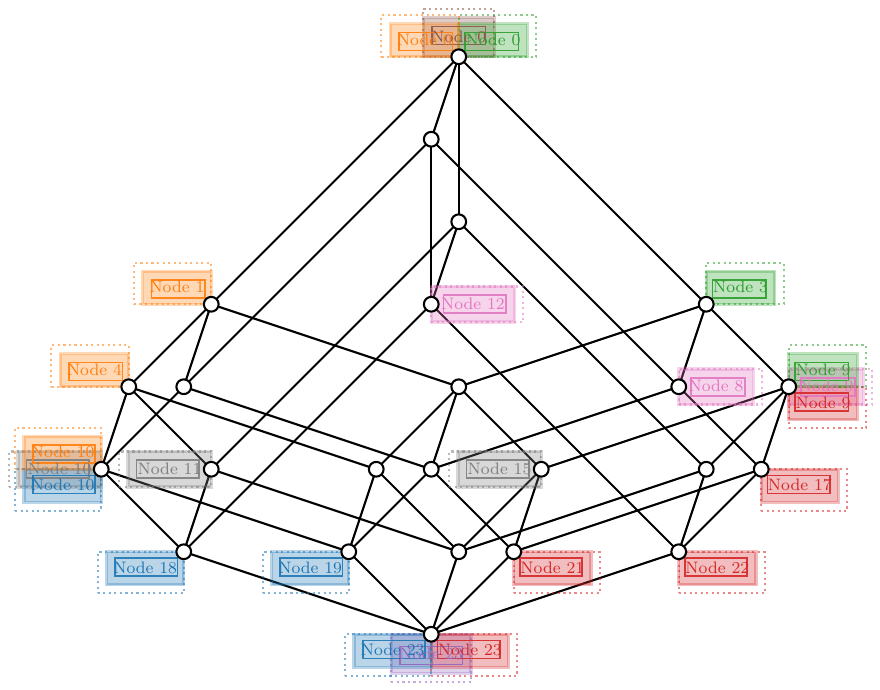}
    \caption{Candidates filtered by intersecting edges (left) and neighbor direction (right).}
    \label{fig:filtered_second}
\end{figure}

Given the local environment $G_v' \subseteq G'$, which is the subgraph consisting of all nodes and edges bounding the faces $F'_v$ of $G'$ that are incident to $v$, we further refine the candidate set $\Lambda$ based on the local geometry of the planarized graph $G'$. Firstly, the node proximity filter eliminates label candidates whose expanded bounding boxes $\RR_{\varepsilon}$ contain any nodes of $G_v'$, which prevents labels from overlapping with vertices:
\[
\Phi_{\text{node}}(\ell) = \{ \lambda \in \Phi_{\text{out}}(\ell) \mid \forall u \in V(G_v') \setminus \{ v \} \colon u \not\in \RR_{\varepsilon}(\lambda) \}.
\]
Furthermore, the edge intersection filter ensures that the label's ink does not overlap with the graph's edges, excluding any candidate whose minimal bounding box $\RR$ intersects an edge in the local neighborhood:
\[
\Phi_{\text{edge}}(\ell) = \{ \lambda \in \Phi_{\text{node}}(\ell) \mid \forall e \in E(G_v') \colon e \cap \RR(\lambda) = \emptyset \}.
\]
Finally, the angular sector filter exploits the direction of incident edges to prioritize candidates in free geometric sectors. Rather than calculating the exact intersection geometries, this filter partitions the plane around $v$ into four $\pi/2$-sectors corresponding to the Cartesian quadrants $\{Q_{TL}, Q_{TR}, Q_{BL}, Q_{BR}\}$. For a node $v \in V$, let $N(v) = \{u \in V \mid v \prec u \text{ or } u \prec v\}$ denote the set of incident neighbors in $G$. We define a set of prohibited anchor positions $\AA_{\not\angle}$, to identify candidates that stand in angular opposition to the available whitespace:
\[
\AA_{\not\angle} = \bigcup_{Q \in \{Q_{\text{TL}}, Q_{\text{TR}}, Q_{\text{BL}}, Q_{\text{BR}}\}} \bigcup \{ \operatorname{restr}(Q) \mid \exists u \in N(v) : \operatorname{atan2}(y_u - y_v, x_u - x_v) \in \operatorname{sector}(Q) \},
\]
where the restriction mapping $\operatorname{restr}(Q)$ identifies the candidates to be pruned for each occupied sector:
\[
\operatorname{restr}(Q) = 
\begin{cases} 
    \{\text{B}, \text{R}, \text{BR}\} & \text{if } Q = Q_{\text{TL}} \\
    \{\text{B}, \text{L}, \text{BL}\} & \text{if } Q = Q_{\text{TR}} \\
    \{\text{T}, \text{R}, \text{TR}\} & \text{if } Q = Q_{\text{BL}} \\
    \{\text{T}, \text{L}, \text{TL}\} & \text{if } Q = Q_{\text{BR}}.
\end{cases}
\]
To ensure that the filter does not eliminate all viable positions for a label in high-density regions, we define the angular sector filter such that it only prunes the candidate set if a non-empty subset of valid positions remains:
\[
\Phi_{\text{ang}}(\ell) =
\begin{cases} 
    \{ \lambda \in \Phi_{\text{edge}}(\ell) \mid \operatorname{anc}(\lambda) \not\in \AA_{\not\angle}(\ell) \} & \text{if } \exists \lambda \in \Phi_{\text{edge}}(\ell) \colon \operatorname{anc}(\lambda) \not\in \AA_{\not\angle}(\ell) \\
    \Phi_{\text{edge}}(\ell) & \text{otherwise.}
\end{cases}
\]
These local geometry filters ensure that the algorithm first satisfies hard constraints (no overlaps with nodes or edges) and then uses the angular heuristic to reduce the candidate set, before getting to the more computationally expensive steps of label placement.

%%%%%%%%%%%%%%%%%%%%%%%%%%%%%%%%%%%%%%%%%%%%%%%%%%%%%%%%%%%%%%%%%%%%%%%%%%%%%%%%
\subsection{Readability-driven conflict resolution}
\label{subsec:readability_filter}
%%%%%%%%%%%%%%%%%%%%%%%%%%%%%%%%%%%%%%%%%%%%%%%%%%%%%%%%%%%%%%%%%%%%%%%%%%%%%%%%
To select a maximal set of non-overlapping labels, we adapted the hybrid algorithm of Wagner and Wolff~\cite{Wagner1998}. While we retain their original combination of rule-based logic~\cite{Wagner2001} and Kakoulis-Tollis matching heuristics~\cite{Kakoulis1998}, we introduce a readability-driven tie-breaker to further optimize placement.

In the first phase, we resolve conflicts, defined as the intersection of two candidate bounding boxes $\RR_\delta(\lambda_a)$ and $\RR_\delta(\lambda_b)$ from different labels $\ell_i$ and $\ell_j$, using the original selection rules of the hybrid algorithm, adjusted to account for multiple independent labels per node:

\begin{itemize}
    \item \textbf{Safe Selection:} If a label candidate $\lambda_a \in \Lambda_i$ has zero active conflicts with any other candidate $\lambda_b \in \Lambda \setminus \Lambda_i$, it is immediately selected.
    \item \textbf{Mutual Dependency:} If a candidate $\lambda_a \in \Lambda_i$ conflicts only with $\lambda_b \in \Lambda_j$, and conversely, $\lambda_d \in \Lambda_j$ conflicts only with $\lambda_c \in \Lambda_i$, the conflict is resolved by selecting a non-conflicting pair, such as $\{ \lambda_a, \lambda_d \}$.
    \item \textbf{Clique Resolution:} If a label $\ell$ has only one remaining candidate $\lambda_a$ and all its conflicting candidates form a clique, \ie, they all conflict with each other, then $\lambda_a$ is selected.
\end{itemize}

If multiple candidates fulfill these rules, we deviate from the original algorithm's~\cite{Wagner2001} first-come, first-served selection by prioritizing candidates based on graph-specific visual criteria. Although this step adds a layer of computation, the performance impact is negligible due to the initial filtering of candidates, which for most graphs leaves only a few remaining conflicts. For labels assigned to interior nodes, we prioritize candidates closest to the centroid of the largest incident face $f' \in F'(v)$, thereby anchoring the label within the largest available white space. For labels assigned to nodes on the boundary walk $\partial f'_\infty$, we prioritize the candidate minimizing the weighted score $\CC(\lambda) = |t - 0.5| - \alpha d_\perp$, where $t = \frac{(p_{\lambda} - p_{\text{prev}}) \cdot w}{\|w\|^2}$ represents the projection of the candidate center onto the chord vector $w = p_{\text{next}} - p_{\text{prev}}$ between the closest labeled neighbors along the boundary walk $p_{\text{next}}$ and $p_{\text{prev}}$. This balances centering along the boundary segment against visual clearance, by incorporating the weighted perpendicular distance $\alpha d_\perp$.

\begin{figure}[t]
    \centering
    \includegraphics[width=0.45\textwidth]{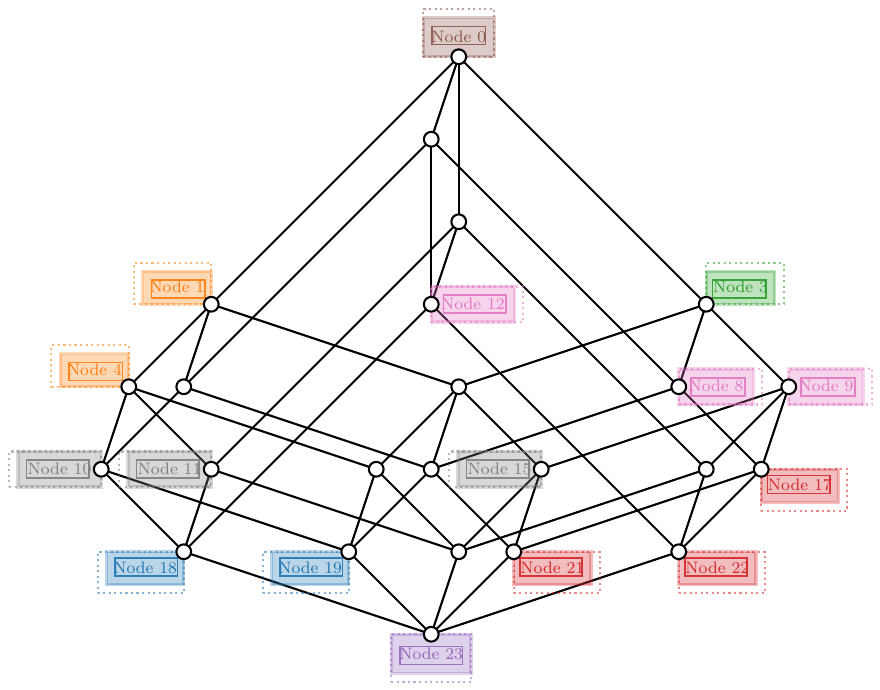}
    \caption{Conflict resolution via hybrid algorithm with a readability-driven tie-breaker.}
    \label{fig:hybrid}
\end{figure}

In the second phase, we resolve remaining conflicts using the matching heuristic of Kakoulis and Tollis~\cite{Kakoulis1998}, modified with an anchor priority tie-breaker that prioritizes top-left placements, motivated by Western reading patterns. A conflict graph $G_{\text{conf}} = (C, E)$ is defined where an edge $(\lambda_i, \lambda_j) \in E$ represents intersecting candidates from distinct labels. To decompose dependencies, the candidate possessing the maximum degree $\Delta(G)$ within each connected component is iteratively eliminated. To maintain node coverage, essential candidates $\lambda$, which are the final remaining option for any label $\ell$, where $|C_{\lambda}| = 1$, are protected from elimination; in such cases, the candidate with the next highest degree is removed instead. This recursive pruning continues until every component is reduced to a clique $K \subseteq C$. Finally, the selection is modeled as a maximum-cardinality bipartite matching on a graph $H = (\Lambda_{\neq} \cup K, E')$, where $\Lambda_{\neq}$ is the set of labels with unresolved conflicts. Solving this matching, with the tie-breaker ensuring a deterministic choice among equivalent candidates, yields a near-optimal placement that maximizes label count while accounting for graph-specific criteria (Figure~\ref{fig:hybrid}).

%%%%%%%%%%%%%%%%%%%%%%%%%%%%%%%%%%%%%%%%%%%%%%%%%%%%%%%%%%%%%%%%%%%%%%%%%%%%%%%%
\subsection{Bounded overflow labels}
\label{subsec:bounded_overflow}
\begin{figure}[t]
    \centering
    \includegraphics[width=0.45\textwidth]{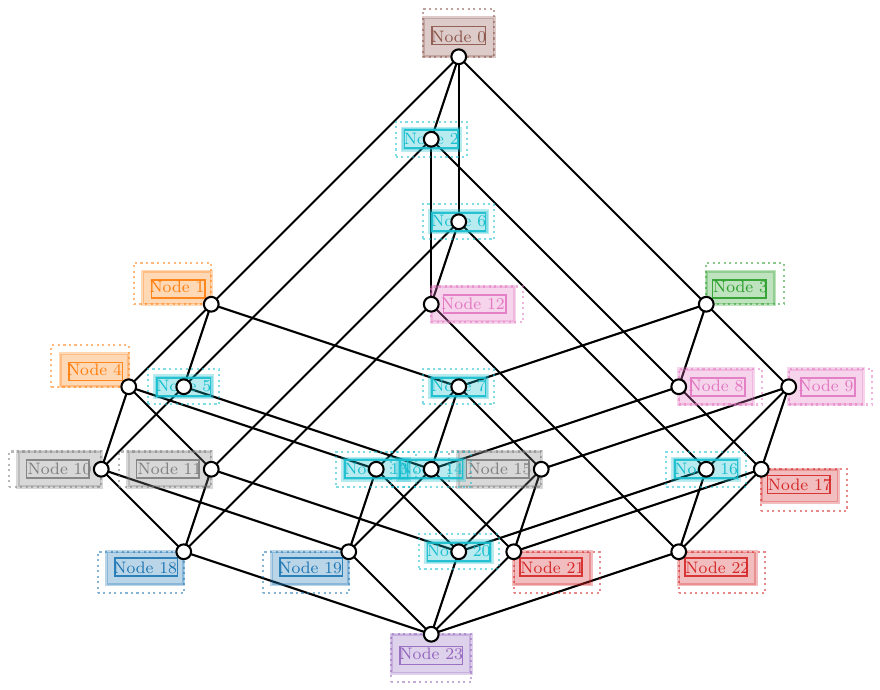}
    \hfill
    \includegraphics[width=0.45\textwidth]{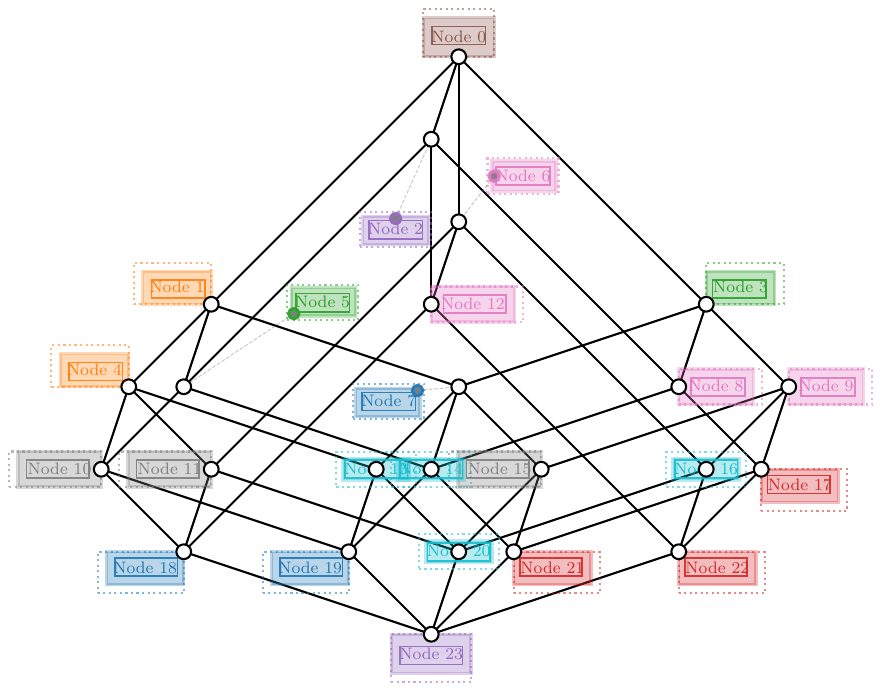}
    \caption{Initial overflow candidates $\Lambda^\times$ (left) and positioned bounded overflow labels (right).}
    \label{fig:bounded_overflow}
\end{figure}

%%%%%%%%%%%%%%%%%%%%%%%%%%%%%%%%%%%%%%%%%%%%%%%%%%%%%%%%%%%%%%%%%%%%%%%%%%%%%%%%
For unplaced labels $\LL^\times = \LL \setminus \LL^*$ after conflict resolution, we introduce bounded overflow labels (Figure~\ref{fig:bounded_overflow}) situated within the faces $F'$ of the planarized graph and connected to its respective node $v \in V$ via a binding line $\sigma$. To preserve visual clarity, we impose a topological constraint that $\sigma$ may intersect at most one edge $e \in E$. We define the valid search space $\SS$ through a topological visibility mask, where a face $f'$ is a candidate if it is either incident to $v$ or shares a boundary edge $e'$ with an incident face, such that the search region is formed by the union of incident faces and the visibility stencils $S_{v, f'}$ created by the intersection of non-incident faces $f'$ with the infinite wedge $W(v, e')$ subtended by $v$:
\[
\SS(\ell) = \left\{ \lambda \in \left( (f' \cap W(v, e')) \setminus \bigcup_{\lambda_j \in \Lambda^*} \RR_{\delta}(\lambda_j) \right) \ominus \RR_{\epsilon}(\ell) \mid \|\mathbf{a}^* - \mathbf{v}\|_2 < \textrm{dist}(\mathbf{v}, \partial f'_\infty) \right\}.
\]
The boundary-aware pruning factor $\|\mathbf{a}^* - \mathbf{v}\|_2 < \textrm{dist}(\mathbf{v}, \partial f'_\infty)$ further ensures that labels are placed more naturally by requiring the distance of the chosen anchor $a^*$ to be closer to the node $v$ than the distance of $v$ to the concave hull $\partial f'_\infty$.

To minimize crossings, we process faces in descending area order, prioritizing unplaced labels $\ell \in \LL^\times$ incident to the face boundary $\partial f'$. This ensures that most overflow labels maintain zero edge crossings. If no incident nodes remain, we select the nearest valid node from an adjacent face, permitting exactly one crossing via the shared edge $e'$. The optimal label center $\lambda^*$ is determined by minimizing the Euclidean distance to the anchor node within the search space $\SS(\ell)$, while the binding line $\sigma$ is anchored at the best valid anchor $a^*$:
\[
\lambda^* = \argmin_{\lambda \in \SS(\ell)} \|\mathbf{\lambda} - \mathbf{v}\|_2, \quad a^* = \argmin_{a \in \AA_{\text{valid}}(\lambda^*)} \|\sigma_{a, v} \|_2.
\]
An anchor $a$ is valid if its binding line $\sigma$ avoids all graph nodes $v \in V$ the minimal bounding rectangles $\RR$ of already placed labels $\LL^*$ and the MBR of its own label $\RR(\lambda^*)$. We define the set of valid anchors $\AA_{\text{valid}}$ for a candidate $\lambda$ as:
\[
\AA_{\text{valid}}(\lambda) = \left \{ a \in \AA \mid \sigma \cap V = \emptyset \text{ and } \sigma \cap \left (\RR(\lambda) \cup \bigcup_{\lambda_j \in \Lambda^*} \RR(\lambda_j) \right ) = \emptyset \right \}.
\]

%%%%%%%%%%%%%%%%%%%%%%%%%%%%%%%%%%%%%%%%%%%%%%%%%%%%%%%%%%%%%%%%%%%%%%%%%%%%%%%%
\subsection{Unbounded overflow labels}
\label{subsec:unbounded_overflow}
%%%%%%%%%%%%%%%%%%%%%%%%%%%%%%%%%%%%%%%%%%%%%%%%%%%%%%%%%%%%%%%%%%%%%%%%%%%%%%%%
To handle the remaining labels $\LL^\times$ that are not placed at their respective node and cannot fit within the bounded faces of $G'$, we migrate them to the infinite exterior $\R^2 \setminus \Gamma$ as overflow labels. We discretize this space into a grid $\PP$ and apply a vectorized boundary-proximity filter to prune the search space to a feasible ``ribbon'' surrounding the graph. Let the boundary $\partial f'_\infty$ be defined by a set of segments $S = \{ s_1, s_2, \ldots, s_n \}$. For each grid center $p \in \PP$, we calculate the distance to the nearest point $p'$ on the boundary. For a segment with endpoints $v_1, v_2$ the projection $t$ is:
\[
t = \operatorname{clamp} \left( \frac{(\mathbf{p} - \mathbf{v}_1) \cdot (\mathbf{v}_2 - \mathbf{v}_1)}{\|\mathbf{v}_2 - \mathbf{v}_1\|^2}, 0, 1 \right).
\]
The distance $d(\mathbf{p}, S) = \|\mathbf{p} - (\mathbf{v}_1 + t(\mathbf{v}_2 - \mathbf{v}_1)) \|$ allows us to discard grid cells where $d < \epsilon$. Vectorizing this filter transforms the global search over $\PP$ into a localized optimization within the ribbon, which keeps the process computationally efficient while ensuring overflow labels stay close to $\Gamma$.

Given the grid points with at least one valid anchor $\AA_{\text{valid}} \neq \emptyset$ after the geometric pruning, we rank the label candidates by calculating a cost function $\CC(p) = \CC_{\text{fix}} + \CC_{\text{var}}$.

The unified fixed cost $\CC_{\text{fix}}(\lambda)$ aggregates the geometric relationship between the grid center $p$, the respective node $v$, and the graph boundary $\partial f'_\infty$:
\[
\CC_{\text{fix}}(\lambda) = \underbrace{w_{\theta} \Delta\theta + w_{\partial} d(a^*, \partial f'_\infty)}_{\text{Global Context}} + \underbrace{w_{\sigma} \|\sigma\| + w_{\phi} [1 - \cos \phi]}_{\text{Local Connectivity}}
\]
\begin{itemize}
    \item \textbf{Angular Preservation ($w_{\theta} \Delta\theta$):} Penalizes angular deviation $\Delta\theta = \text{acos} \left( \frac{\mathbf{p} \cdot \mathbf{v}}{\|\mathbf{p}\| \|\mathbf{v}\|} \right)$ from the node's original radial vector, which ensures labels expand outward without losing their relative positioning.
    
    \item \textbf{Boundary Proximity ($w_{\partial} d$):} Keeps unbounded overflow labels tucked against the graph by minimizing the distance $d(a^*, \partial f'_\infty) = \min_{u \in \partial f'_\infty} \|a^* - u\|$ between the selected anchor point $a^*$ and the boundary of the concave hull $\partial f'_\infty$.
    
    \item \textbf{Anchor Alignment ($w_{\phi} [1 - \cos \phi]$):} Penalizes the angular deviation between the binding line $\sigma = a^* - v$ and the label's internal radial axis $p - a^*$. By minimizing the cost $1 - \cos \phi$, this term favors connections on the side facing the graph.
    
    \item \textbf{Binding Displacement ($w_{\sigma} \|\sigma\|$):} Constraints the length of the binding line $\sigma$, which serves as a proximity check to keep the anchor point $a^*$ tightly linked to its source node, thus preventing visual decoupling.
\end{itemize}

\begin{figure}[t]
    \centering
    \includegraphics[width=0.45\textwidth]{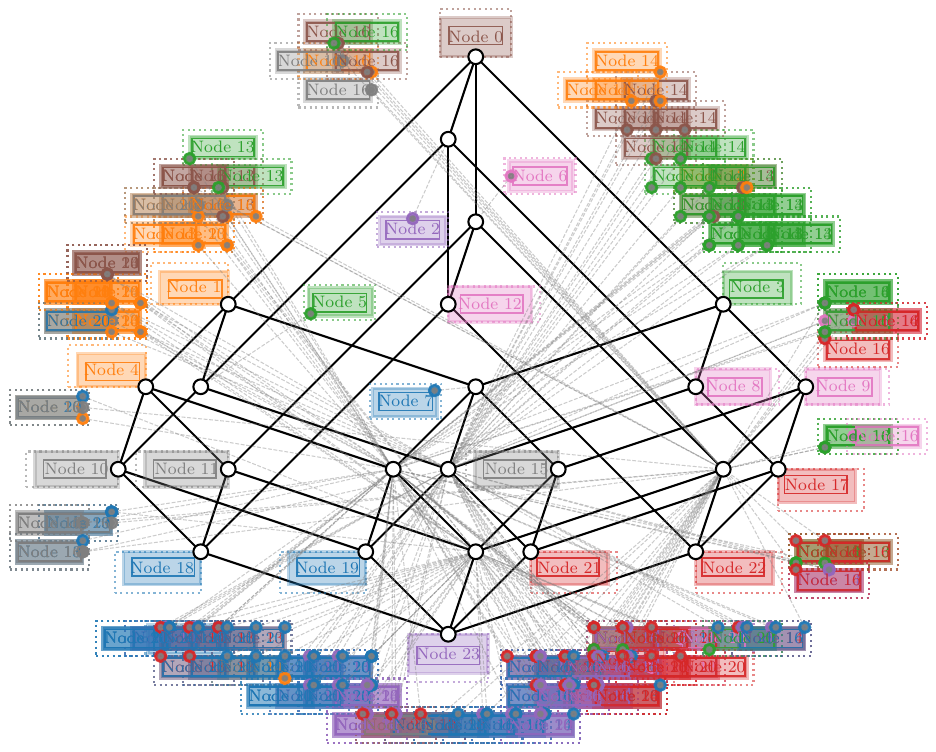}
    \hfill
    \includegraphics[width=0.45\textwidth]{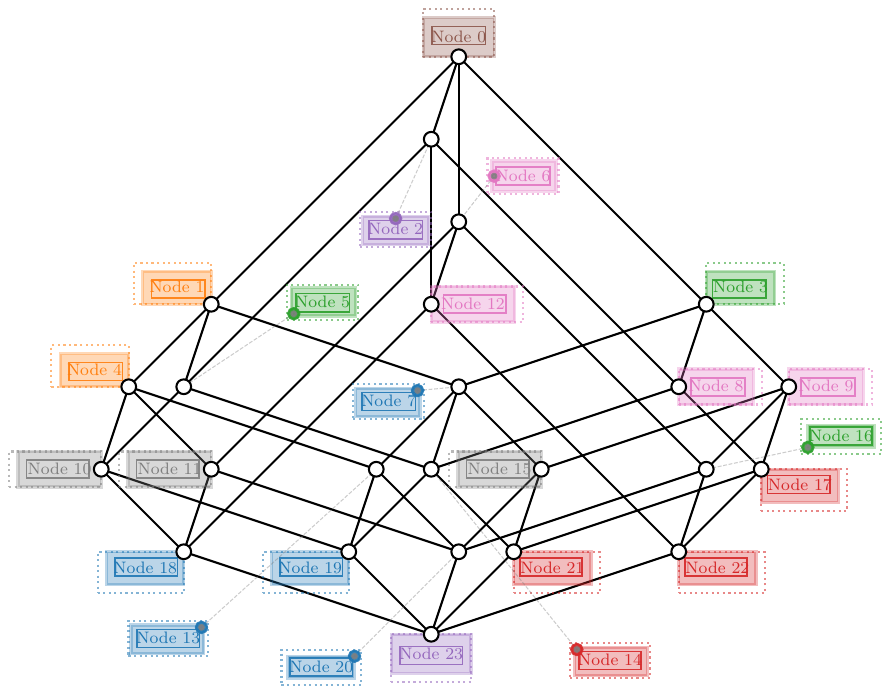}
    \caption{All unbounded grid candidates (left) and positioned overflow labels (right).}
    \label{fig:unbounded_overflow}
\end{figure}

The non-linear cost $\CC_{\text{var}}$ aims to iteratively resolve spatial conflicts. In each iteration $k$, the solver evaluates structural violations between the active selection and the candidate pools to refine the global cost matrix $M$:
\[
M_{i,j}^{(k+1)} = M_{i,j}^{(k)} + \Pi(\lambda_{i,j}) \cdot \gamma^k,
\]
where $\Pi(\lambda_{i,j})$ quantifies the cumulative penalty arising when assigning label $\lambda_i$ to candidate $\lambda_j$. This value is derived by summing conflicts with the active label candidates $\Lambda^*_m$ currently assigned to any other node:
\[
\Pi(\lambda_{i,j}) = \sum_{m \neq i} \left( w_{\lambda \lambda} |\lambda_{i,j} \cap \Lambda^*_m| + w_{\sigma \lambda} |\sigma_{i,j} \cap \Lambda^*_m| + w_{\sigma \sigma} |\sigma_{i,j} \cap \sigma_m| \right).
\]
By applying a multiplier $\gamma > 1$ to progressively inflate specific matrix entries, we force the Hungarian algorithm~\cite{Kuhn1955} to explore a broader candidate space. Since this feedback loop does not always converge on a singular, stable state, we monitor the results over a predefined number of iterations; if convergence is not reached, we keep the configuration that yielded the lowest global cost during the run. This approach allows us to resolve spatial overlaps by intentionally trading local placement precision for global topological clarity (Figure~\ref{fig:unbounded_overflow}).

%%%%%%%%%%%%%%%%%%%%%%%%%%%%%%%%%%%%%%%%%%%%%%%%%%%%%%%%%%%%%%%%%%%%%%%%%%%%%%%%
\subsection{Force-based refinement}
\label{subsec:force}
\begin{figure}[t]
    \centering
    \includegraphics[width=0.45\textwidth]{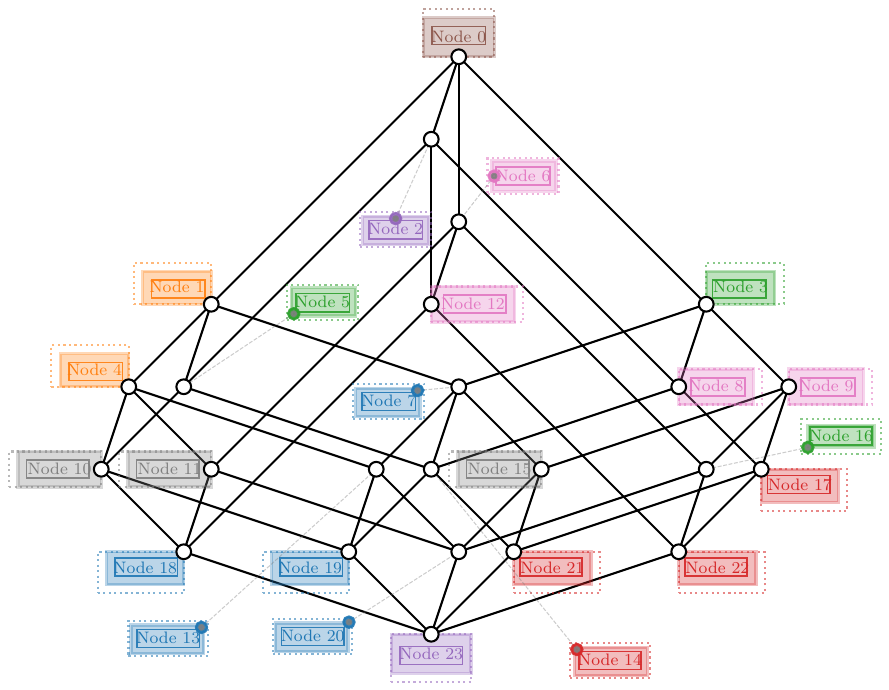}
    \caption{Final layout after force refinement.}
    \label{fig:force_refined}
\end{figure}

%%%%%%%%%%%%%%%%%%%%%%%%%%%%%%%%%%%%%%%%%%%%%%%%%%%%%%%%%%%%%%%%%%%%%%%%%%%%%%%%
As a final step, we refine the positions of unbounded overflow labels using a force-directed model that treats labels as dynamic particles within a physical system (Figure~\ref{fig:force_refined}). Let each unbounded overflow candidate $\lambda_i \in \Lambda^\times$ be represented by its center $c_ i \in \R^2$ and its minimal bounding rectangle $\RR(\lambda_i)$. We define the total force $\mathbf{F}(\lambda_i)$ acting on a label candidate as the vector sum of the individual components that push or pull the label to an optimized position.

To improve spacing and prevent label overlaps, the repulsion force $\mathbf{F}_{\text{rep}}$ shifts the label tangentially relative to the drawing centroid $\mathbf{C}_{\Gamma}$. Let $\mathbf{\hat{r}}_i = \frac{c_i - \mathbf{C}_{\Gamma}}{\|c_i - \mathbf{C}_{\Gamma}\|}$ be the radial unit vector, and $\mathbf{\hat{t}}_i = [-\hat{r}_{iy}, \hat{r}_{ix}]^T$ be the perpendicular tangent vector. We define the repulsive force as:
\[
\mathbf{F}_{\text{rep}}(\lambda_i) = \sum_{\lambda_j \in \Lambda^*} \max(0, \tau_{\text{rep}} - \|\mathbf{p}_i - \mathbf{p}_j\|) \cdot \operatorname{sgn}(\mathbf{\hat{r}}_i \times \mathbf{\hat{v}}_{ij}) \cdot \mathbf{\hat{t}}_i,
\]
which identifies the nearest points $p_i \in \partial \lambda_i$ and $p_j \in \partial \lambda_j$, sliding the label along the arc $\mathbf{\hat{t}}_i$ whenever their distance falls below the threshold $\tau_{\text{rep}}$. The direction of this shift is determined by the cross product between $\mathbf{\hat{r}}_i$ and the placed label's radial direction $\mathbf{\hat{v}}_{ij} = \frac{\mathbf{p}_j - \mathbf{C}_{\Gamma}}{\|\mathbf{p}_j - \mathbf{C}_{\Gamma}\|}.$

To achieve a uniform distribution of labels around the drawing $\Gamma$, $\mathbf{F}_{\text{tan}}$ applies a lateral shift along the same tangential axis $\mathbf{\hat{t}}_i$. This adjustment improves label spacing while preserving their original distance from the centroid:
\[
\mathbf{F}_{\text{tan}}(\lambda_i) = \sum_{j \neq i} \max(0, \tau_{\text{gap}} - \text{dist}(\RR_i, \RR_j)) \cdot \operatorname{sgn}(\mathbf{\hat{r}}_i \times \mathbf{\hat{r}}_j) \cdot \mathbf{\hat{t}}_i.
\]
The force moves the label clockwise or counterclockwise based on the relative angular position of neighboring labels $\mathbf{\hat{r}}_j$, ensuring a balanced distribution around the drawing.

The spring force $\mathbf{F}_{\text{spring}}$ constrains the distance from the anchor point $a_i$ and the intersection point $p_\partial$ of the binder $\sigma$ and the boundary of the concave hull $\partial f'_\infty$. Using a quadratic penalty for stronger corrective pull at greater distances, the force is:
\[
\mathbf{F}_{\text{spring}}(\lambda_i) = -k (\|\mathbf{a}_i - \mathbf{p}_\partial\| - \sigma^*)^2 \cdot \mathbf{\hat{u}}_{\text{gap}},
\]
where $a_i$ is the anchor point, $\sigma^*$ is the target binder length, and $\mathbf{\hat{u}}_{\text{gap}}$ is the unit vector pointing from the drawing boundary toward the label. 

To prevent the binder segment $\sigma_i$ from intersecting obstacles $\OO_j$, such as nodes $V$ or the ink of labels $\RR_j \neq \RR_i$, $\mathbf{F}_{\text{steer}}$ applies a tangential correction: 
\[
\mathbf{F}_{\text{steer}}(\lambda_i) = \sum_{o_j \in \OO_j} \max(0, \tau_{\text{bind}} - \text{dist}(\sigma_i, o_j)) \cdot \operatorname{sgn}(\mathbf{\hat{r}}_i \times \mathbf{\hat{v}}_{obs}) \cdot \mathbf{\hat{t}}_i.
\]
If the gap between the binder $\sigma_i$ and an obstacle falls below the clearance threshold $\tau_{\text{bind}}$, the label is shifted tangentially. This creates a ``swinging'' motion that allows the binder to pivot around the obstacle $\mathbf{\hat{v}}_{obs}$ while preserving the label's radial distance.

%%%%%%%%%%%%%%%%%%%%%%%%%%%%%%%%%%%%%%%%%%%%%%%%%%%%%%%%%%%%%%%%%%%%%%%%%%%%%%%%
\section{Labeling concept lattices}
\label{sec:cl_algorithm}
%%%%%%%%%%%%%%%%%%%%%%%%%%%%%%%%%%%%%%%%%%%%%%%%%%%%%%%%%%%%%%%%%%%%%%%%%%%%%%%%
Formal Concept Analysis~\cite{GanterFCA2024} is a framework for human centered data analysis that derives a hierarchy of formal concepts from a set of objects and their shared attributes (Figure~\ref{fig:driving_concepts_cxt}). The derived concept lattice is visually represented as a line diagram, where nodes represent formal concepts and edges represent the covering relation between them. To maintain readability, labels are categorized into two types: intent labels (attributes) and extent labels (objects) that are visually separated by their vertical position relative to the respective node and their font style.

A \emph{formal context} is a triple $\K := \GMI$, consisting of a set $G$ of objects, a set $M$ of attributes, and an incidence relation $I \subseteq G \times M$. Within this framework, $g \relI m$ ($(g,m) \in \relI$) indicates that object $g$ has attribute $m$. For a context $\GMI$, we define the derivation operators $(\cdot)'$ for any $A \subseteq G$ and $B \subseteq M$ as:
\[
A' := \{ m \in M \mid \forall g \in A \colon (g, m) \in I \},
\]
representing the attributes common to all objects in $A$, and dually:
\[
B' := \{ g \in G \mid \forall m \in B \colon (g, m) \in I \},
\]
the set of objects posessing all attributes in $B$.

A \emph{formal concept} is a pair $(A, B)$ with $A \subseteq G$ and $B \subseteq M$ such that $A' = B$ and $B' = A$. Here, $A$ is called the the \emph{extent} and $B$ the \emph{intent} of the concept. The set of all formal concepts $\BGMI$ is ordered by the following relation:
\[
(A_1, B_1) \leq (A_2, B_2) \iff A_1 \subseteq A_2.
\]
The resulting \emph{concept lattice} $\BVGMI = (\BGMI, \leq)$ is a complete lattice.

Following standard conventions in the FCA community, we use the following reduced labeling to maintain visual clarity. Each attribute $m \in M$ is attached only to the attribute concept $\mu m = \{ \{m\}', \{m\}'' \}$, the most general concept that includes $m$ in its intent. Similarly, each object $g \in G$ is attached to the object concept $\gamma g = \{ \{g\}'', \{g\}' \}$, the unique, most specific concept containing $g$ in its extent. Intent labels are positioned above their respective node in italics, while extent labels are placed below in an upright font.

\begin{figure}[t]
    \centering
    \begin{cxt}[b]
        \att{\rotatebox{90}{$Dl_+$}}
        \att{\rotatebox{90}{$M_+$}}
        \att{\rotatebox{90}{$R_+$}}
        \att{\rotatebox{90}{$De_+$}}
        \att{\rotatebox{90}{$E_+$}}
        \att{\rotatebox{90}{$Dl_{++}$}}
        \att{\rotatebox{90}{$De_{++}$}}
        \att{\rotatebox{90}{$Cl$}}
        \att{\rotatebox{90}{$M_-$}}
        \att{\rotatebox{90}{$Su$}}
        \att{\rotatebox{90}{$E_-$}}
        \att{\rotatebox{90}{$M_{++}$}}
        \att{\rotatebox{90}{$Cm$}}
        \att{\rotatebox{90}{$De_-$}}
        \att{\rotatebox{90}{$Ch$}}
        \att{\rotatebox{90}{$Su/n$}}
        \att{\rotatebox{90}{$Cvl$}}
        \att{\rotatebox{90}{$R_{++}$}}
        \att{\rotatebox{90}{$E_{++}$}}
        \att{\rotatebox{90}{$Dl_-$}}
        \att{\rotatebox{90}{$Sn$}}
        \att{\rotatebox{90}{$E_{--}$}}
        \att{\rotatebox{90}{$M_{--}$}}
        \att{\rotatebox{90}{$R_{--}$}}
        \att{\rotatebox{90}{$So$}}
        \obj{xxx.x....x.xxx...........}{Standard}
        \obj{x.xxxxx.x.....xx.........}{All-Wheel}
        \obj{.xxxx..x.x......xxxx.....}{Front-wheel}
        \obj{x.xx.xxxx.x.........xxx..}{Mid-engine}
        \obj{xx.x.xxx..x............xx}{Rear-wheel}
    \end{cxt}
    \caption{Formal context of driving concepts~\cite{GanterFCA2024}.}
    \label{fig:driving_concepts_cxt}
\end{figure}

%%%%%%%%%%%%%%%%%%%%%%%%%%%%%%%%%%%%%%%%%%%%%%%%%%%%%%%%%%%%%%%%%%%%%%%%%%%%%%%%
\subsection{Pre-filter the 8-position model}
\label{subsec:pre_filter}
\begin{figure}[t]
    \centering
    \includegraphics[width=0.45\textwidth]{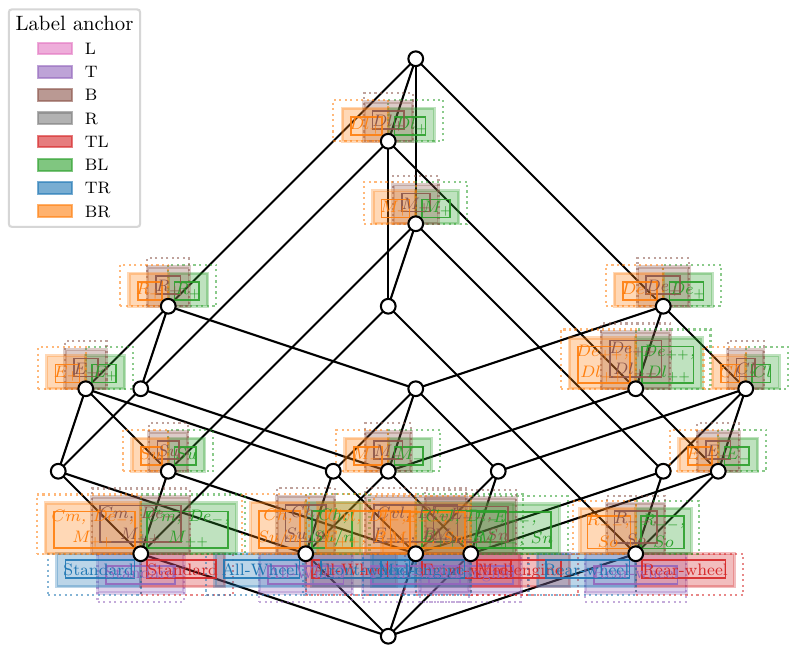}
    \hfill
    \includegraphics[width=0.45\textwidth]{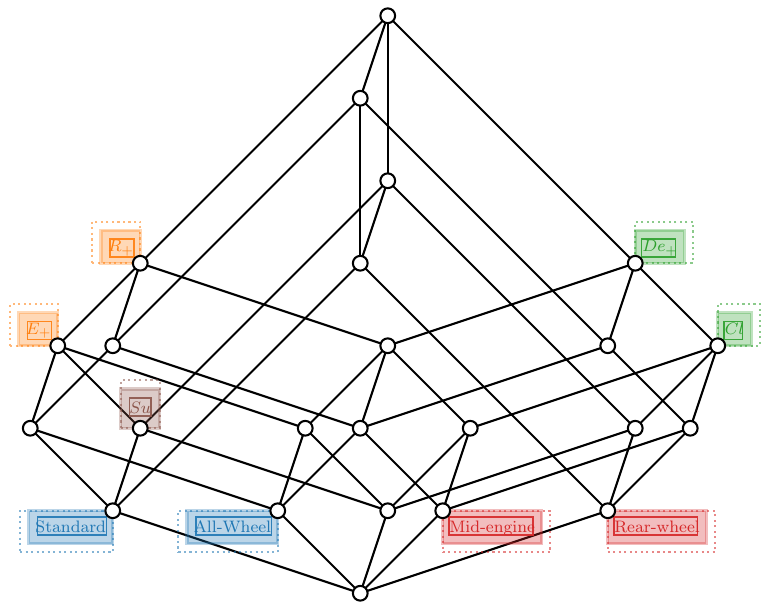}
    \caption{Candidate labels for intent and extent labels before (left) and after filtering (right).}
    \label{fig:fca_candidates}
\end{figure}

%%%%%%%%%%%%%%%%%%%%%%%%%%%%%%%%%%%%%%%%%%%%%%%%%%%%%%%%%%%%%%%%%%%%%%%%%%%%%%%%
To align with the standard FCA conventions for label placement, we prune the initial set of candidates $\Lambda$ generated by the 8-position model based on the label conventions before applying the local geometric filters (Section~\ref{subsec:local_filter}). We restrict intent labels to candidates with an anchor on the bottom side of the bounding box and extent labels to candidates with an anchor on the top side:
\[
\Phi_{\text{type}}(\ell) = 
\begin{cases}
    \{\lambda_{\text{BL}}, \lambda_{\text{B}}, \lambda_{\text{BR}}\} & \text{if } \ell \in \LL_{\text{intent}} \\
    \{\lambda_{\text{TL}}, \lambda_{\text{T}}, \lambda_{\text{TR}}\} & \text{if } \ell \in \LL_{\text{extent}}.
\end{cases}
\]
This pre-filtering step ensures that the subsequent filtering pipeline operates only on valid candidates, which reduces the local search space for each node (Figure~\ref{fig:fca_candidates}). For some labels, this may result in an empty set of candidates. Such labels are treated as overflow labels in the subsequent steps.

%%%%%%%%%%%%%%%%%%%%%%%%%%%%%%%%%%%%%%%%%%%%%%%%%%%%%%%%%%%%%%%%%%%%%%%%%%%%%%%%
\subsection{Prune search space for bounded overflow labels}
\label{subsec:fca_bounded}
\begin{figure}[t]
    \centering
    \includegraphics[width=0.45\textwidth]{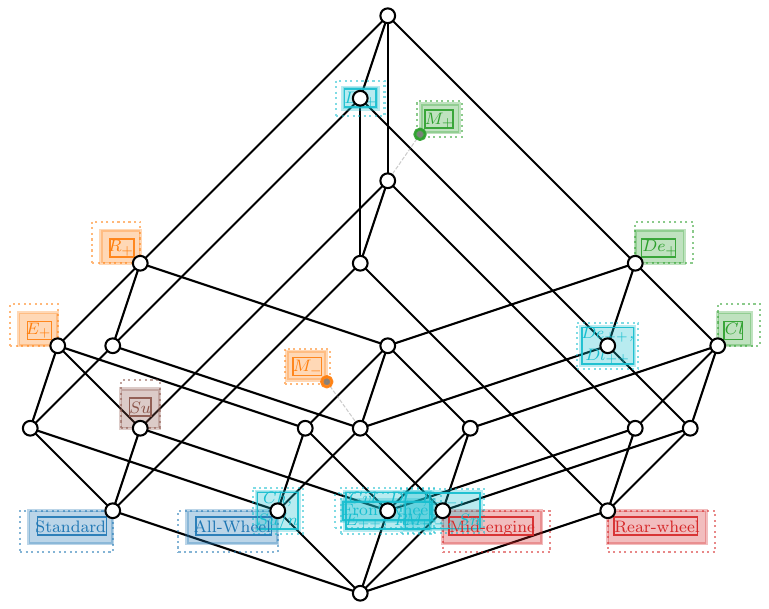}
    \caption{Bounded overflow placement for intent and extent labels.}
    \label{fig:fca_bounded}
\end{figure}

%%%%%%%%%%%%%%%%%%%%%%%%%%%%%%%%%%%%%%%%%%%%%%%%%%%%%%%%%%%%%%%%%%%%%%%%%%%%%%%%
To ensure that the bounded overflow labels $\ell \in \LL \setminus \LL^*$ within the drawing $\Gamma$ align with FCA conventions, we prune the search space $\SS(\ell)$ based on the respective node's vertical position. Therefore, we restrict the search space so that the label's bounding box $\RR(\ell)$ remains entirely within the semantically correct half-plane $\HH_v$ relative to $v_y$: 
\[
\SS_{\text{type}}(\ell) = \SS(\ell) \cap
\begin{cases}
    \{ p \mid y_{\text{min}}(\RR(\lambda_p)) > v_y \} & \text{if } \ell \in \LL_{\text{intent}} \\
    \{ p \mid y_{\text{max}}(\RR(\lambda_p)) < v_y \} & \text{if } \ell \in \LL_{\text{extent}}.
\end{cases}
\]
While ensuring visual consistency, we also optimize the search process by filtering out faces in the planarized graph that have insufficient area within the desired half-plane (Figure~\ref{fig:fca_bounded}).

\begin{figure}[t]
    \centering
    \includegraphics[width=0.45\textwidth]{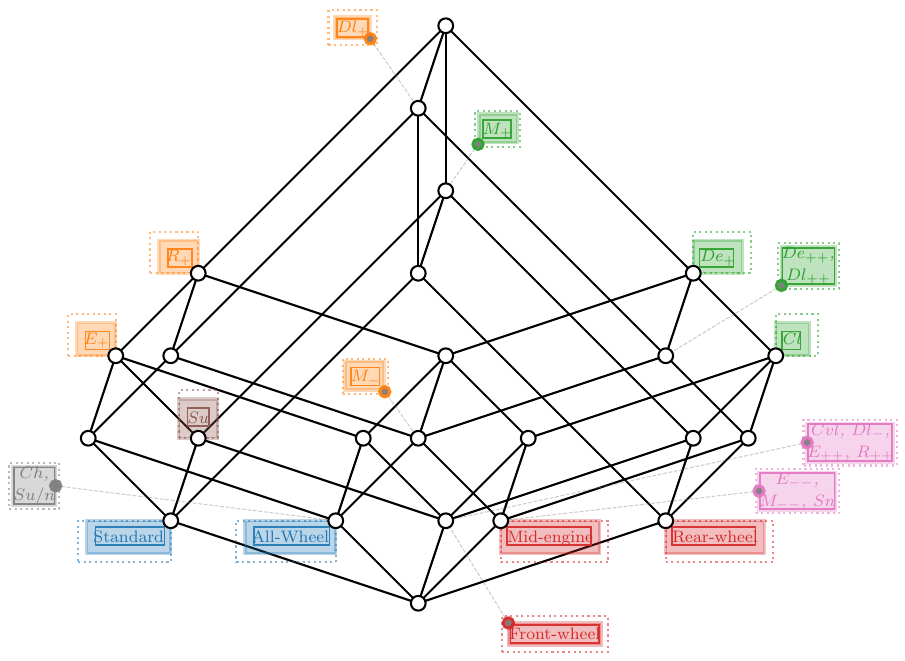}
    \hfill
    \includegraphics[width=0.45\textwidth]{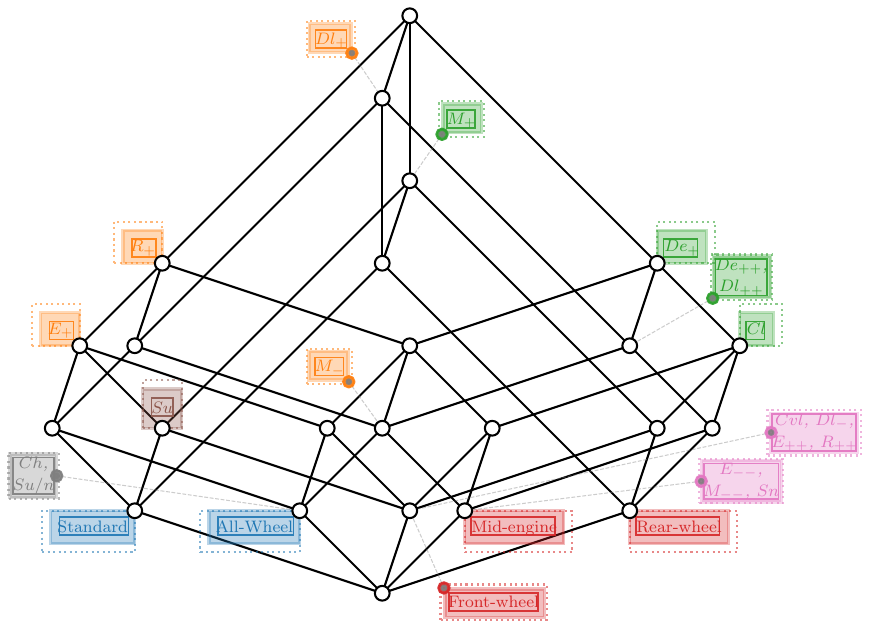}
    \caption{Unbounded overflow placement (left) and the force-refined placement (right).}
    \label{fig:fca_unbounded}
\end{figure}

%%%%%%%%%%%%%%%%%%%%%%%%%%%%%%%%%%%%%%%%%%%%%%%%%%%%%%%%%%%%%%%%%%%%%%%%%%%%%%%%
\subsection{Extended cost function for unbounded overflow labels}
\label{subsec:fca_unbounded}
%%%%%%%%%%%%%%%%%%%%%%%%%%%%%%%%%%%%%%%%%%%%%%%%%%%%%%%%%%%%%%%%%%%%%%%%%%%%%%%%
To better balance the FCA conventions with practical feasibility, we have intergrated a directional penalty $\CC_{\text{hp}}$ into the fixed cost $\CC_{\text{fix}}(p)$. Rather than pruning the search space, this approach applies a heavy weight to any grid center $p$ where the label $\RR(\lambda_p)$ lies outside of the preferred half-plane $\HH_v$. By treating these boundaries as soft preferences rather than hard constraints, we enable the Hungarian algorithm to always find a valid global mapping by bypassing the directional rules in crowded areas.

%%%%%%%%%%%%%%%%%%%%%%%%%%%%%%%%%%%%%%%%%%%%%%%%%%%%%%%%%%%%%%%%%%%%%%%%%%%%%%%%
\subsection{Extended force model for refining unbounded overflow labels}
\label{subsec:fca_force}
%%%%%%%%%%%%%%%%%%%%%%%%%%%%%%%%%%%%%%%%%%%%%%%%%%%%%%%%%%%%%%%%%%%%%%%%%%%%%%%%
Finally, we introduce a half-plane restorative force $\mathbf{F}_{\text{hp}}(\lambda_i)$ to pull labels back to their proper half-plane if they drift during the optimization process (Figure~\ref{fig:fca_unbounded}). This force acts tangentially, effectively ``sliding'' a label along its current orbit until it respects the node's vertical boundary. A violation is triggered whenever the bottom edge of an intent label drops below the node $v_y$, or the top edge of an extent label rises above it. To determine the most efficient correction, we use the global up-vector $\mathbf{\hat{j}} = [0, 1]^T$ and a type-dependent multiplier $d_i$ (set to $1$ for intent and $-1$ for extent). The force is calculated as:
\[
\mathbf{F}_{\text{hp}}(\lambda_i) = w_{\text{hp}} \cdot (\delta_i + \epsilon) \cdot \operatorname{sgn} \left((\mathbf{\hat{r}}_i \times \mathbf{\hat{j}}) \cdot d_i \right) \cdot \mathbf{\hat{t}}_i.
\]
Here, $\delta_i$ denotes the vertical violation relative to the threshold $v_y$, a small buffer $\epsilon$ detects labels that are close to the threshold, and the cross product of the radial vector $\mathbf{\hat{r}}_i$ and the up-vector $\mathbf{\hat{j}}$ determines the shortest path to tangentially push the label back into its designated half-plane.

\begin{figure}[t]
    \centering
    \includegraphics[width=0.48\textwidth]{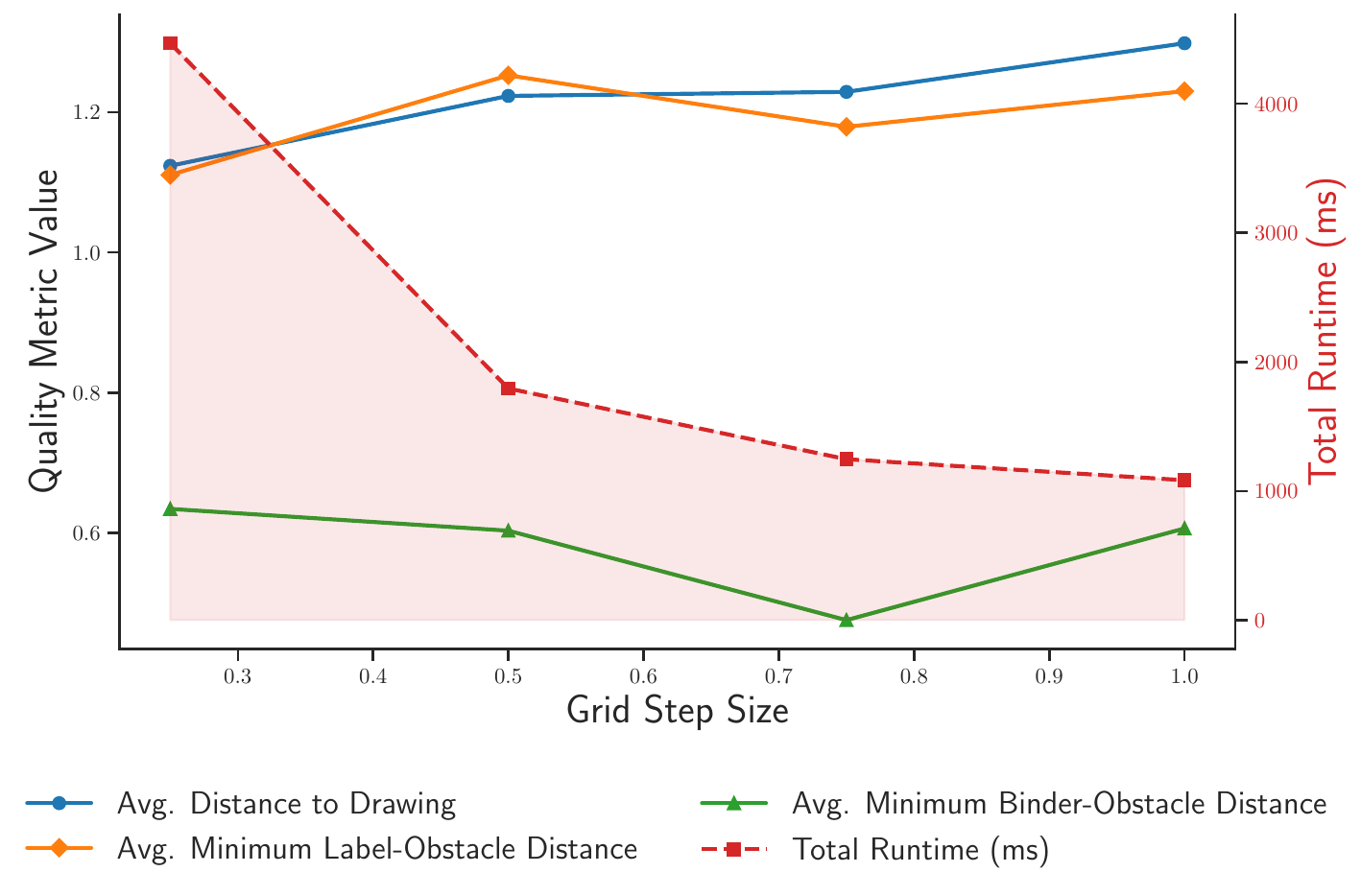}%
    \includegraphics[width=0.48\textwidth]{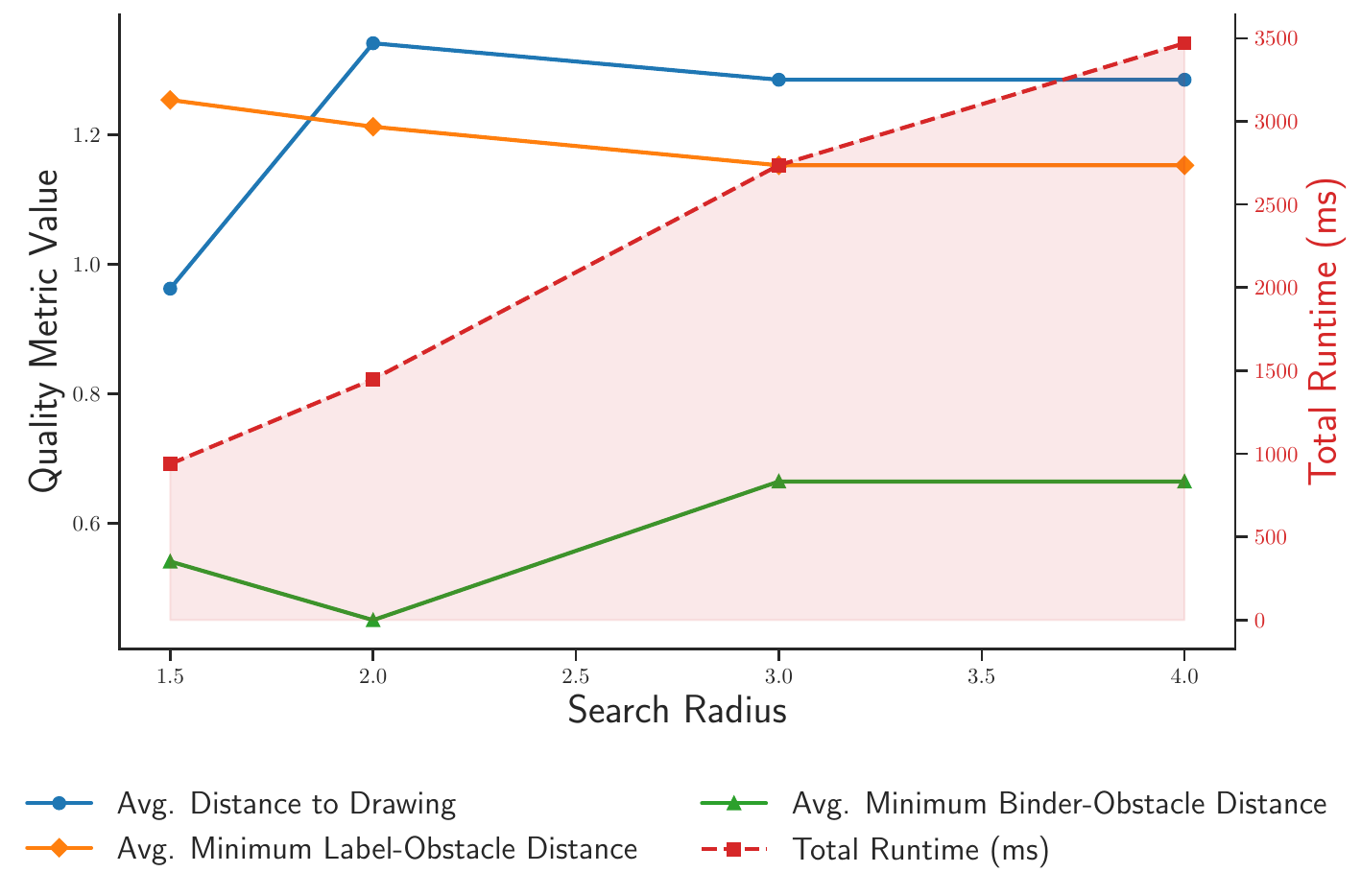}%
    \vspace{1em}
    \includegraphics[width=0.48\textwidth]{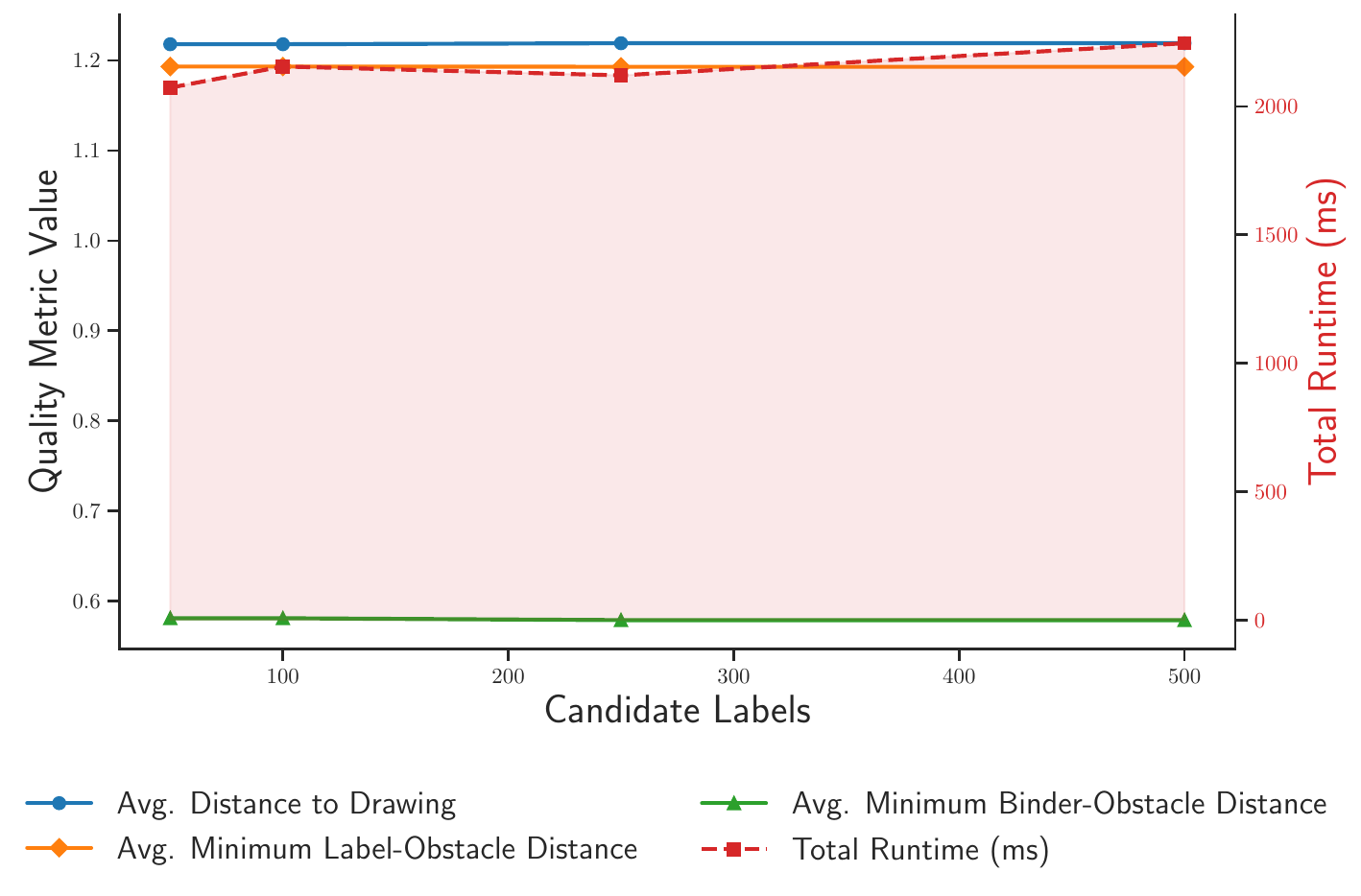}%
    \includegraphics[width=0.48\textwidth]{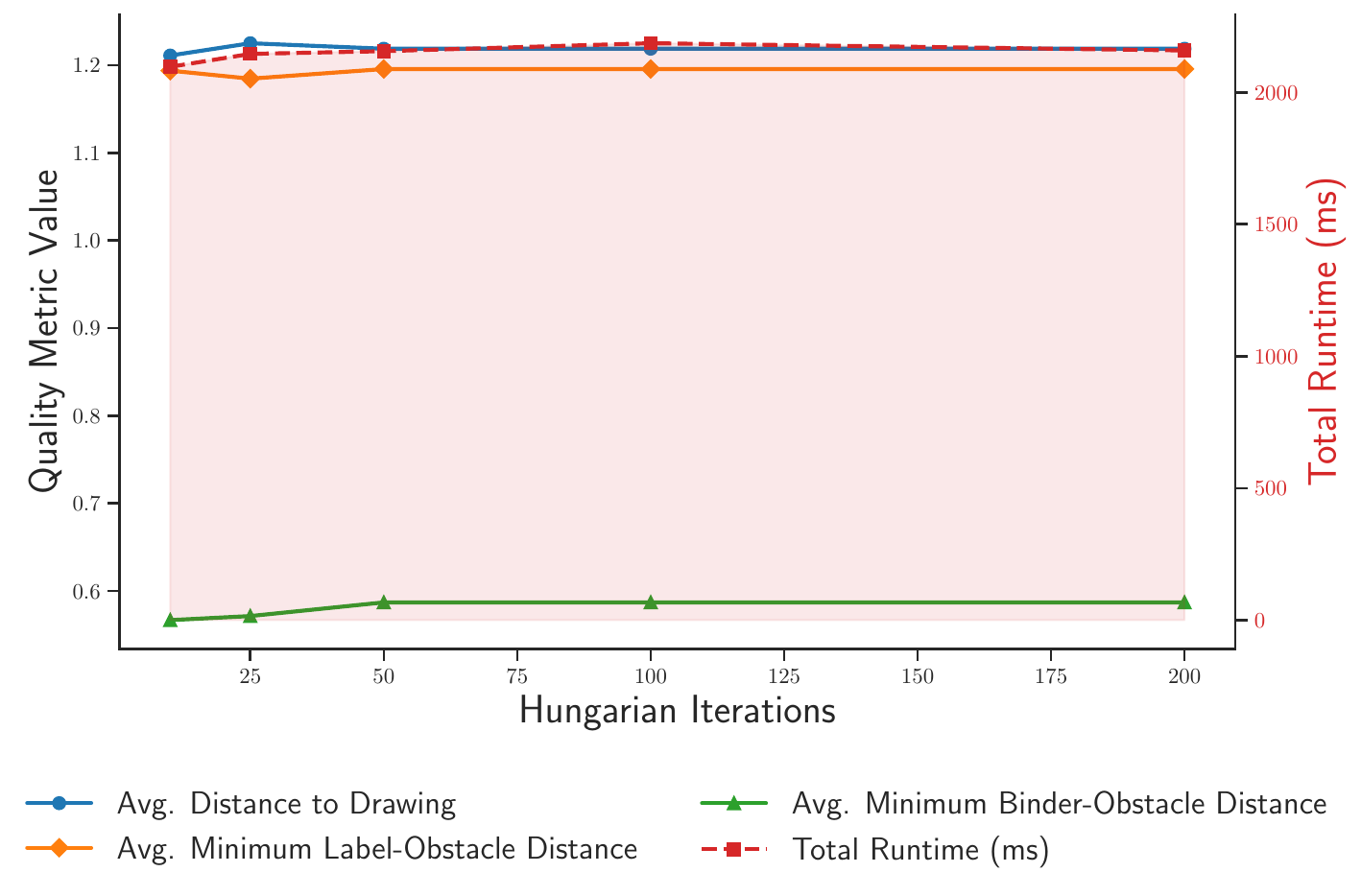}%
    \caption{Trade-off analysis of quality metrics vs. total runtime by tunable parameters.}
    \label{fig:trade_off}
\end{figure}

\begin{figure}[t]
    \centering
    \hspace*{\fill}
    $\vcenter{\hbox{\raisebox{3ex}{\includegraphics[width=0.45\textwidth]{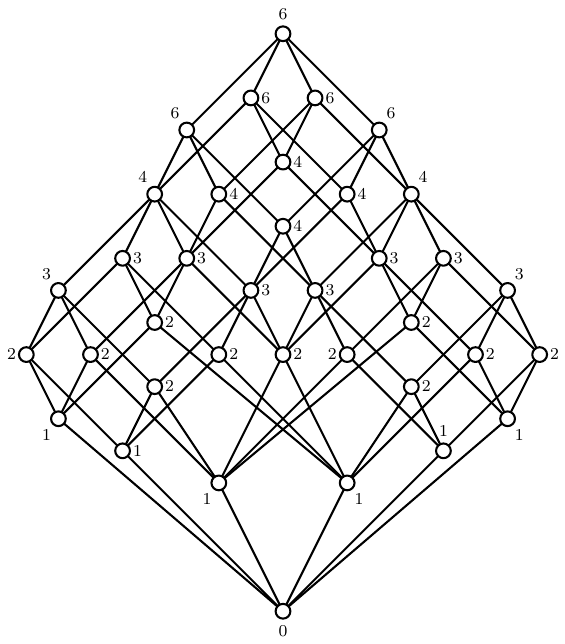}}}}$%
    $\vcenter{\hbox{\includegraphics[width=0.55\textwidth]{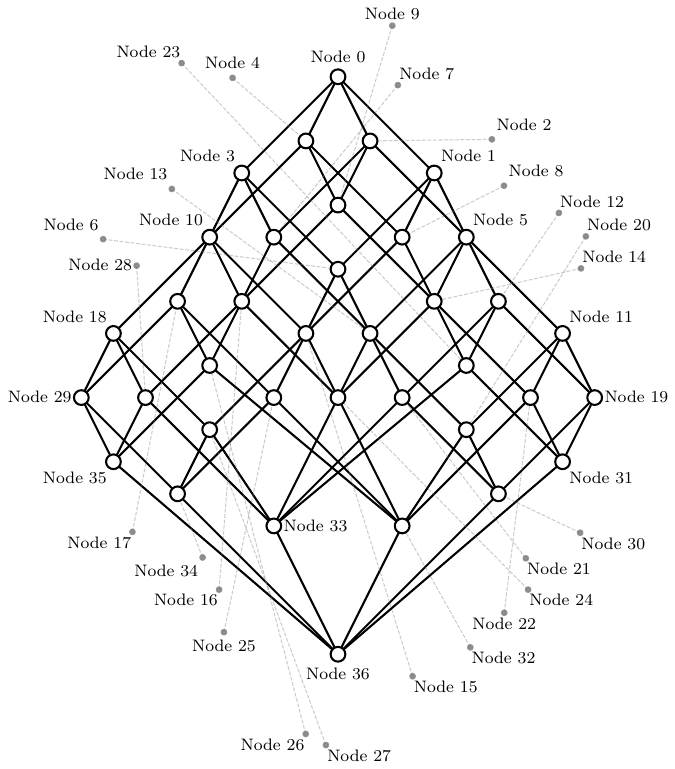}}}$%
    \hspace*{\fill}
    \caption{Labeling the lattice of the convex-ordinal scale with join-rises (left) and indices (right).}
    \label{fig:convex-ordinal}
\end{figure}

%%%%%%%%%%%%%%%%%%%%%%%%%%%%%%%%%%%%%%%%%%%%%%%%%%%%%%%%%%%%%%%%%%%%%%%%%%%%%%%%
\section{Evaluation}
\label{sec:evaluation}
%%%%%%%%%%%%%%%%%%%%%%%%%%%%%%%%%%%%%%%%%%%%%%%%%%%%%%%%%%%%%%%%%%%%%%%%%%%%%%%%
To evaluate the trade-off between computational efficiency and output quality (Figure~\ref{fig:trade_off}), we measure runtime across several tunable variables: the top-$k$ candidates and maximum iterations for the Hungarian solver, the grid step size, the pruning threshold for the grid search space, and the iteration depth of force refinement. Therefore, we consider standard FCA graphs~\cite{Duerrschnabel2020} with up to 30 nodes and synthetic label sizes varying from 1 to 3 lines, differing in width. To assess the resulting layout quality, we consider three spatial metrics: the average distance of unbounded overflow labels to the drawing, the mean minimum distance of labels to obstacles, and the mean minimum distance of binding lines to obstacles.

Investigating the runtime shows that the initial rendering overhead of approximately 300\,ms is dominated by the \LaTeX\ engine for supporting math mode, which can be reduced to a few milliseconds for plain text labels. Furthermore, the primary bottleneck is constructing the grid of unbounded overflow candidates, which scales as $O(n \cdot |P| \cdot \log n)$; here, spatial indexing via the STR-based R-tree reduces intersection tests to logarithmic time~\cite{Leutenegger1997} relative to the grid density $|P| \propto (d_{\text{max}} / \rho)^2$. While the subsequent Hungarian solver has a theoretical complexity of $O(n^3)$~\cite{Kuhn1955}, it remains performant in practice. Finally, a force-based refinement step resolves interactions between labels and obstacles $\OO$. Using the same efficient spatial indexing, this step also scales logarithmically with the number of obstacles; however, the process typically converges within 25 to 100 iterations, leading to practical efficiency.

A sensitivity analysis of the tunable parameters shows that the performance of our algorithm is mainly affected by the grid construction phase, which can be significantly accelerated by increasing the grid step size $\rho = 0.5$ and reducing the maximum search distance $d_{\text{max}} = 2.0$. Since the force-based refinement step effectively optimizes label positions, this pruning of the search space leads to minimal impact on accuracy. Furthermore, the Hungarian solver plateaus quickly; limiting it to the top 50 candidates and 10 iterations improves efficiency without loss of quality. While we set a high iteration ceiling ($10^4$) for the force-based model, the system typically converges within 25 to 100 iterations.

Investigating well-known concept lattices (such as those shown in Figure~\ref{fig:evaluation_real_world}) that are labeled based on the proposed default parameterization reveals that our approach effectively manages real-world examples from the standard book on Formal Concept Analysis~\cite{GanterFCA2024}. In edge cases, such as the dense convex-ordinal scale lattice (Figure~\ref{fig:convex-ordinal}), we observe that most numerical data, like the join-rises~\cite{Abdulla2025}, which naturally require smaller labels, can be easily placed at their respective nodes. In contrast, larger labels, \ie, the previous indexing of nodes using labels of the format ``Node \textit{index}'' that require more space, are placed further away due to the large number of labels that need to be placed around the drawing. In these cases, our approach prioritizes non-overlapping positions, which leads to longer binding lines and more frequent intersections; however, the clear association between labels and their respective nodes is maintained. Users can further reduce the number of intersections by considering more top-$k$ candidates for the Hungarian solver or decreasing the label size.

Thus, scalability is naturally bounded by the available space for label placement, since label count and size compete for the same finite area. In practice, this is rarely an issue, since the FCA community switches to alternative representations, such as nested line diagrams~\cite{GanterFCA2024}, once a lattice grows too large for a plain line diagram to remain readable.  

%%%%%%%%%%%%%%%%%%%%%%%%%%%%%%%%%%%%%%%%%%%%%%%%%%%%%%%%%%%%%%%%%%%%%%%%%%%%%%%%
\section{Conclusion}
\label{sec:conclusion}
%%%%%%%%%%%%%%%%%%%%%%%%%%%%%%%%%%%%%%%%%%%%%%%%%%%%%%%%%%%%%%%%%%%%%%%%%%%%%%%%
Our findings indicate that our algorithm effectively manages most real-world scenarios. Routine tasks, such as node indexing or assigning numerical data, are handled with a minimal need for overflow labels. While larger labels naturally necessitate more overflow labels, the algorithm is able to find a valid position within an adjacent face or the nearest whitespace of the graph surrounding. However, in challenging cases such as dense graphs, a large number of labels to be placed, or large labels, the limited whitespace naturally impacts label proximity. Accordingly, prioritizing non-overlapping positions leads to longer binding lines and more frequent intersections. Despite these trade-offs, our approach maintains a clear association between labels and their respective nodes.

Future research should explore a transition from the rigid fixed-position anchoring toward sliding models~\cite{Kreveld1999}, allowing for more granular label placement and improved anchoring for overflow labels via sliding binding anchors. Furthermore, the integration of dynamic rendering, such as more flexible line breaks or using concave hulls instead of MBRs, might improve the positioning of complex labels. To further enhance the placement of overflow labels, considering B\'ezier curves~\cite{Fink2012} instead of straight binding lines should improve the final positions due to more flexibility; however, the shift requires a refactored grid computation strategy to maintain pruning efficiency. Finally, extending the labeling approach to account for the structural constraints of nested line diagrams presents a promising direction for future work.

%%
%% Bibliography
%%
\bibliography{bibliography}

\appendix

\end{document}

%% file: figures/algorithm.tex
\definecolor{tabblue}{HTML}{1F77B4}
\definecolor{taborange}{HTML}{FF7F0E}
\definecolor{tabgreen}{HTML}{2CA02C}
\definecolor{tabred}{HTML}{D62728}
\definecolor{tabpurple}{HTML}{9467BD}
\definecolor{tabbrown}{HTML}{8C564B}
\definecolor{tabpink}{HTML}{E377C2}
\definecolor{tabgray}{HTML}{7F7F7F}
\definecolor{tabolive}{HTML}{BCBD22}
\definecolor{tabcyan}{HTML}{17BECF}
\colorlet{lpBgIO}{tabgray!20}
\colorlet{lpBgSpatial}{tabblue!20}
\colorlet{lpBgFixed}{taborange!20}
\colorlet{lpBgFilt}{tabgreen!20}
\colorlet{lpBgBounded}{tabbrown!20}
\colorlet{lpBgUnbounded}{tabpurple!20}
\colorlet{lpBgOpt}{tabpink!20}
\tikzset{
  lp@base/.style={
    rectangle, rounded corners=3pt, 
    minimum width=28mm, minimum height=6.5mm, 
    font=\footnotesize\sffamily, text=white,
    inner sep=3pt, align=center, line width=0.35pt
  },
  lp@input/.style={
    lp@base,
    fill=tabgray, draw=tabgray!70!black,
    minimum width=60mm, rounded corners=5pt
  },
  lp@output/.style={
    lp@base,
    fill=tabgray, draw=tabgray!70!black,
    minimum width=28mm, rounded corners=5pt
  },
  lp@proc/.style= {lp@base, fill=tabblue, draw=tabblue!60!black},
  lp@fix/.style= {lp@base, fill=taborange, draw=taborange!60!black},
  lp@filt/.style= {lp@base, fill=tabgreen, draw=tabgreen!60!black},
  lp@bounded/.style= {lp@base, fill=tabbrown, draw=tabbrown!60!black},
  lp@unbounded/.style= {lp@base, fill=tabpurple, draw=tabpurple!60!black},
  lp@opt/.style= {lp@base, fill=tabpink, draw=tabpink!60!black},
  lp@arr/.style={
    -{Stealth[length=4pt, width=3.5pt]},
    line width=0.5pt, draw=gray!80!black,
  },
  lp@grp/.style={
    rectangle, rounded corners=3pt,
    line width=0.4pt, dashed,
    inner xsep=4pt, inner ysep=6pt,
  },
  lp@caption/.style={
    font=\small\sffamily, text=tabgray!80!black,
    align=center,
  }
}
\resizebox{\linewidth}{!}{%
    \begin{tikzpicture}[node distance=4mm and 4mm]
        \node[lp@input] (input) {Line Diagram $\Gamma$, Labels $\LL$ \\ (\ref{sec:problem})};
        \node[lp@proc, right=of input] (partition) {Partition \\ (\ref{subsec:label})};
        \node[lp@proc, right=of partition] (render) {Render \\ (\ref{subsec:label})};
        \node[lp@fix, right=of render] (cand) {Initial Candidates \\ (\ref{subsec:fixed_position})};
        \node[lp@filt, below=12mm of cand] (fout) {{\large $\Phi_{\mathrm{out}}$} \\ (\ref{subsec:local_filter})};
        \node[lp@filt, left=of fout] (fnode) {{\large $\Phi_{\mathrm{node}}$} \\ (\ref{subsec:local_filter})};
        \node[lp@filt, left=of fnode] (fedge) {{\large $\Phi_{\mathrm{edge}}$} \\ (\ref{subsec:local_filter})};
        \node[lp@filt, left=of fedge] (fang) {{\large $\Phi_{\mathrm{ang}}$} \\ (\ref{subsec:local_filter})};
        \node[lp@filt, left=of fang] (conflict) {Conflict Resolution \\ (\ref{subsec:readability_filter})};
        \node[lp@bounded, below=12mm of conflict] (overflow) {Bounded Overflow \\ (\ref{subsec:bounded_overflow})};
        \node[lp@unbounded, right=of overflow] (grid) {Grid Candidates \\ (\ref{subsec:unbounded_overflow})};
        \node[lp@unbounded, right=of grid] (hung) {Hungarian Solver \\ (\ref{subsec:unbounded_overflow})};
        \node[lp@opt, right=of hung] (force) {Force Refinement \\ (\ref{subsec:force})};
        \node[lp@output, right=of force] (output) {Labeled Line Diagram \\ (Fig. \ref{fig:force_refined})};
        \begin{scope}[on background layer]
            \node[lp@grp, draw=tabgray!60, fill=lpBgIO, fit=(input),
                    label={[font=\small\sffamily\bfseries\color{tabgray!80!black}]above:Input}] {};
            \node[lp@grp, draw=tabblue!60, fill=lpBgSpatial, fit=(partition)(render),
                    label={[font=\small\sffamily\bfseries\color{tabblue!80!black}]above:Spatial Quantification}] {};
            \node[lp@grp, draw=taborange!60, fill=lpBgFixed, fit=(cand),
                    label={[font=\small\sffamily\bfseries\color{taborange!80!black}]above:Fixed Positions}] {};
            \node[lp@grp, draw=tabgreen!60, fill=lpBgFilt, fit=(fout)(fnode)(fedge)(fang)(conflict),
                    label={[font=\small\sffamily\bfseries\color{tabgreen!80!black}]above:Filter Fixed Positions}] {};
            \node[lp@grp, draw=tabbrown!60, fill=lpBgBounded, fit=(overflow),
                    label={[font=\small\sffamily\bfseries\color{tabbrown!80!black}]above:Bounded Overflow}] {};
            \node[lp@grp, draw=tabpurple!60, fill=lpBgUnbounded, fit=(hung)(grid),
                    label={[font=\small\sffamily\bfseries\color{tabpurple!80!black}]above:Unbounded Overflow}] {};
            \node[lp@grp, draw=tabpink!60, fill=lpBgOpt, fit=(force),
                    label={[font=\small\sffamily\bfseries\color{tabpink!80!black}]above:Optimization}] {};
            \node[lp@grp, draw=tabgray!60, fill=lpBgIO, fit=(output),
                    label={[font=\small\sffamily\bfseries\color{tabgray!80!black}]above:Output}] {};
        \end{scope}
        \draw[lp@arr] (input.east) -- (partition.west);
        \draw[lp@arr] (partition) -- (render);
        \draw[lp@arr] (render) -- (cand);
        \draw[lp@arr] (cand) -- (fout);
        \draw[lp@arr] (fout) -- (fnode);
        \draw[lp@arr] (fnode) -- (fedge);
        \draw[lp@arr] (fedge) -- (fang);
        \draw[lp@arr] (fang) -- (conflict);
        \draw[lp@arr] (conflict) -- (overflow);
        \draw[lp@arr] (overflow) -- (grid);
        \draw[lp@arr] (grid) -- (hung);
        \draw[lp@arr] (hung) -- (force);
        \draw[lp@arr] (force) -- (output);
    \end{tikzpicture}%
}

%% file: bibliography.bib
@article{Knuth1981,
    author = {Knuth, Donald E. and Plass, Michael F.},
    title = {Breaking paragraphs into lines},
    journal = {Software: Practice and Experience},
    volume = {11},
    number = {11},
    pages = {1119-1184},
    doi = {https://doi.org/10.1002/spe.4380111102},
    url = {https://onlinelibrary.wiley.com/doi/abs/10.1002/spe.4380111102},
    eprint = {https://onlinelibrary.wiley.com/doi/pdf/10.1002/spe.4380111102},
    year = {1981}
}

@inproceedings{Formann1991,
    author = {Formann, Michael and Wagner, Frank},
    title = {A packing problem with applications to lettering of maps},
    year = {1991},
    isbn = {0897914260},
    publisher = {Association for Computing Machinery},
    address = {New York, NY, USA},
    url = {https://doi.org/10.1145/109648.109680},
    doi = {10.1145/109648.109680},
    booktitle = {Proceedings of the Seventh Annual Symposium on Computational Geometry},
    pages = {281–288},
    numpages = {8},
    location = {North Conway, New Hampshire, USA},
    series = {SCG '91}
}

@article{Doddi1998,
    author = {Doddi, Srinivas and Marathe, Madhav and Mirzaian, Andy and Moret, Bernard and Zhu, Binhai},
    year = {1998},
    month = {06},
    pages = {},
    title = {Map Labeling and Its Generalizations},
    journal = {Proceedings of the Annual ACM-SIAM Symposium on Discrete Algorithms}
}

@inproceedings{Stein2008,
    author = {Stein, Thierry and D\'{e}coret, Xavier},
    title = {Dynamic label placement for improved interactive exploration},
    year = {2008},
    isbn = {9781605581507},
    publisher = {Association for Computing Machinery},
    address = {New York, NY, USA},
    url = {https://doi.org/10.1145/1377980.1377986},
    doi = {10.1145/1377980.1377986},
    booktitle = {Proceedings of the 6th International Symposium on Non-Photorealistic Animation and Rendering},
    pages = {15–21},
    numpages = {7},
    location = {Annecy, France},
    series = {NPAR '08}
}

@article{Christensen1995,
    author = {Christensen, Jon and Marks, Joe and Shieber, Stuart},
    title = {An empirical study of algorithms for point-feature label placement},
    year = {1995},
    issue_date = {July 1995},
    publisher = {Association for Computing Machinery},
    address = {New York, NY, USA},
    volume = {14},
    number = {3},
    issn = {0730-0301},
    url = {https://doi.org/10.1145/212332.212334},
    doi = {10.1145/212332.212334},
    journal = {ACM Trans. Graph.},
    month = jul,
    pages = {203–232},
    numpages = {30}
}

@inproceedings{Kakoulis1998,
    author = {Kakoulis, Konstantinos G. and Tollis, Ioannis G.},
    title = {A unified approach to labeling graphical features},
    year = {1998},
    isbn = {0897919734},
    publisher = {Association for Computing Machinery},
    address = {New York, NY, USA},
    url = {https://doi.org/10.1145/276884.276923},
    doi = {10.1145/276884.276923},
    booktitle = {Proceedings of the Fourteenth Annual Symposium on Computational Geometry},
    pages = {347–356},
    numpages = {10},
    location = {Minneapolis, Minnesota, USA},
    series = {SCG '98}
}

@InProceedings{Wagner1998,
    author="Wagner, Frank
    and Wolff, Alexander",
    editor="Whitesides, Sue H.",
    title="A Combinatorial Framework for Map Labeling",
    booktitle="Graph Drawing",
    year="1998",
    publisher="Springer Berlin Heidelberg",
    address="Berlin, Heidelberg",
    pages="316--331",
    isbn="978-3-540-37623-1"
}

@Article{Wagner2001,
    author={Wagner, F.
    and Wolff, A.
    and Kapoor, V.
    and Strijk, T.},
    title={Three Rules Suffice for Good Label Placement},
    journal={Algorithmica},
    year={2001},
    month={Jun},
    day={01},
    volume={30},
    number={2},
    pages={334-349},
    issn={1432-0541},
    doi={10.1007/s00453-001-0009-7},
    url={https://doi.org/10.1007/s00453-001-0009-7}
}

@article{Kreveld1999,
    title = {Point labeling with sliding labels},
    journal = {Computational Geometry},
    volume = {13},
    number = {1},
    pages = {21-47},
    year = {1999},
    issn = {0925-7721},
    doi = {https://doi.org/10.1016/S0925-7721(99)00005-X},
    url = {https://www.sciencedirect.com/science/article/pii/S092577219900005X},
    author = {Marc {van Kreveld} and Tycho Strijk and Alexander Wolff}
}

@article{Freeman1975,
    author = {Freeman, H. and Shapira, R.},
    title = {Determining the minimum-area encasing rectangle for an arbitrary closed curve},
    year = {1975},
    issue_date = {July 1975},
    publisher = {Association for Computing Machinery},
    address = {New York, NY, USA},
    volume = {18},
    number = {7},
    issn = {0001-0782},
    url = {https://doi.org/10.1145/360881.360919},
    doi = {10.1145/360881.360919},
    journal = {Commun. ACM},
    month = jul,
    pages = {409–413},
    numpages = {5}
}

@article{Guttman1984,
    author = {Guttman, Antonin},
    title = {R-trees: a dynamic index structure for spatial searching},
    year = {1984},
    issue_date = {June 1984},
    publisher = {Association for Computing Machinery},
    address = {New York, NY, USA},
    volume = {14},
    number = {2},
    issn = {0163-5808},
    url = {https://doi.org/10.1145/971697.602266},
    doi = {10.1145/971697.602266},
    journal = {SIGMOD Rec.},
    month = jun,
    pages = {47–57},
    numpages = {11}
}

@INPROCEEDINGS{Leutenegger1997,
    author={Leutenegger, S.T. and Lopez, M.A. and Edgington, J.},
    booktitle={Proceedings 13th International Conference on Data Engineering}, 
    title={STR: a simple and efficient algorithm for R-tree packing}, 
    year={1997},
    volume={},
    number={},
    pages={497-506},
    doi={10.1109/ICDE.1997.582015}
}

@article{Orenstein1986,
    author = {Orenstein, Jack A.},
    title = {Spatial query processing in an object-oriented database system},
    year = {1986},
    issue_date = {June 1986},
    publisher = {Association for Computing Machinery},
    address = {New York, NY, USA},
    volume = {15},
    number = {2},
    issn = {0163-5808},
    url = {https://doi.org/10.1145/16856.16886},
    doi = {10.1145/16856.16886},
    journal = {SIGMOD Rec.},
    month = jun,
    pages = {326–336},
    numpages = {11}
}

@article{Brinkhoff1993,
    author = {Brinkhoff, Thomas and Kriegel, Hans-Peter and Seeger, Bernhard},
    title = {Efficient processing of spatial joins using R-trees},
    year = {1993},
    issue_date = {June 1, 1993},
    publisher = {Association for Computing Machinery},
    address = {New York, NY, USA},
    volume = {22},
    number = {2},
    issn = {0163-5808},
    url = {https://doi.org/10.1145/170036.170075},
    doi = {10.1145/170036.170075},
    journal = {SIGMOD Rec.},
    month = jun,
    pages = {237–246},
    numpages = {10}
}

@article{Kuhn1955,
    author = {Kuhn, H. W.},
    title = {The Hungarian method for the assignment problem},
    journal = {Naval Research Logistics Quarterly},
    volume = {2},
    number = {1-2},
    pages = {83-97},
    doi = {https://doi.org/10.1002/nav.3800020109},
    url = {https://onlinelibrary.wiley.com/doi/abs/10.1002/nav.3800020109},
    eprint = {https://onlinelibrary.wiley.com/doi/pdf/10.1002/nav.3800020109},
    year = {1955}
}

@ARTICLE{Fink2012,
    author={Fink, Martin and Haunert, Jan-Henrik and Schulz, André and Spoerhase, Joachim and Wolff, Alexander},
    journal={IEEE Transactions on Visualization and Computer Graphics}, 
    title={Algorithms for Labeling Focus Regions}, 
    year={2012},
    volume={18},
    number={12},
    pages={2583-2592},
    doi={10.1109/TVCG.2012.193}
}

@book{GanterFCA2024,
	title         = {Formal Concept Analysis - Mathematical Foundations},
	subtitle      = {Mathematical Foundations},
	author        = {Bernhard Ganter and Rudolf Wille},
	publisher     = {Springer Cham},
	year          = {2024},
	edition       = {2},
	pages         = {XII, 370},
	isbn          = {978-3-031-63421-5},
	ebookisbn     = {978-3-031-63422-2},
	softcoverisbn = {978-3-031-63424-6},
	address       = {Cham, Switzerland}
}

@ARTICLE{sugiyama1981methods,
	author={Sugiyama, Kozo and Tagawa, Shojiro and Toda, Mitsuhiko},
	journal={IEEE Transactions on Systems, Man, and Cybernetics}, 
	title={Methods for Visual Understanding of Hierarchical System Structures}, 
	year={1981},
	volume={11},
	number={2},
	pages={109-125},
	doi={10.1109/TSMC.1981.4308636}
}

@InProceedings{freese2004automated,
	author="Freese, Ralph",
	editor="Eklund, Peter",
	title="Automated Lattice Drawing",
	booktitle="Concept Lattices",
	year="2004",
	publisher="Springer Berlin Heidelberg",
	address="Berlin, Heidelberg",
	pages="112--127",
	isbn="978-3-540-24651-0"
}

@article{Duerrschnabel2023,
  author       = {Dominik D{\"{u}}rrschnabel and
                  Tom Hanika and
                  Gerd Stumme},
  title        = {Drawing Order Diagrams Through Two-Dimension Extension},
  journal      = {CoRR},
  volume       = {abs/1906.06208},
  year         = {2019},
  url          = {http://arxiv.org/abs/1906.06208},
  eprinttype    = {arXiv},
  eprint       = {1906.06208},
  bibsource    = {dblp computer science bibliography, https://dblp.org}
}

@misc{Noehre2026,
	title={DimFlux: Force-Directed Additive Line Diagrams}, 
	author={Marcel Nöhre and Dominik Dürrschnabel and Bernhard Ganter and Gerd Stumme},
	year={2026},
	eprint={2603.16366},
	archivePrefix={arXiv},
	primaryClass={cs.CG},
	url={https://arxiv.org/abs/2603.16366}, 
}

@InProceedings{Abdulla2025,
    author="Abdulla, Mohammad
    and Hille, Tobias
    and D{\"u}rrschnabel, Dominik
    and Stumme, Gerd",
    editor="Cellier, Peggy
    and Ganter, Bernhard
    and Missaoui, Rokia",
    title="Rises for Measuring Local Distributivity in Lattices",
    booktitle="Conceptual Knowledge Structures",
    year="2025",
    publisher="Springer Nature Switzerland",
    address="Cham",
    pages="392--407",
    isbn="978-3-032-03364-2"
}

@misc{Duerrschnabel2020,
  author = {Dürrschnabel, Dominik and Hanika, Tom and Stumme, Gerd},
  title = {Dataset of User Study for DimDraw},
  month = oct,
  year = 2020,
  publisher = {Zenodo},
  doi = {10.5281/zenodo.4075207},
  url = {https://doi.org/10.5281/zenodo.4075207},
  howpublished = {Zenodo},
  note = {Dataset}
}

@INPROCEEDINGS{Pattison2014,
  author={Pattison, Tim and Weber, Derek and Ceglar, Aaron},
  booktitle={2014 IEEE Pacific Visualization Symposium}, 
  title={Enhancing Layout and Interaction in Formal Concept Analysis}, 
  year={2014},
  volume={},
  number={},
  pages={248-252},
  doi={10.1109/PacificVis.2014.21}
}

@book{Davey2002,
	title = {Introduction to Lattices and Order},
	author = {Davey, B. A. and Priestley, H. A.},
	doi = {10.1017/cbo9780511809088},
	isbn = {9780511809088},
	month = {4},
	publisher = { Cambridge University Press },
	year = {2002},
}

@incollection{Birkhoff1967,
	title = {Lattice theory},
	author = {Birkhoff, Garrett},
	booktitle = {Colloquium Publications},
	edition = {3.},
	publisher = {American Mathematical Society},
	volume = {25},
	year = {1967},
}

@book{Diestel2017,
    author = {Diestel, Reinhard},
    title = {Graph Theory},
    year = {2017},
    isbn = {3662536218},
    publisher = {Springer Publishing Company, Incorporated},
    edition = {5th}
}
